\RequirePackage{snapshot}
\RequirePackage{fix-cm}
\documentclass{svjour3} 

\smartqed
\usepackage{graphicx}
\usepackage{array}
\usepackage{pifont}
\usepackage[export]{adjustbox}
\usepackage{natbib}
\usepackage{caption}
\usepackage{subcaption}
\usepackage{multirow}
\usepackage{rotating}
\usepackage{longtable}
\usepackage{tikz}
\usepackage{soul}
\usepackage{xspace}
\usepackage{tikzsymbols}
\usepackage{enumitem}
\usepackage{amsmath}
\usepackage{url}
\usepackage{hyperref}
\usepackage{slashbox}
\usepackage{amsmath,amssymb,amsfonts}
\usepackage{algorithmic}
\usepackage{textcomp}
\usepackage{xcolor}
\usepackage{booktabs, tabularx}
\usepackage{listings}
\usepackage{makecell}
 \bibpunct{(}{)}{,}{a}{,}{,}

\newcommand{\figref}[1]{Figure~\ref{#1}\xspace}

\newcommand{\tabref}[1]{Table~\ref{#1}\xspace}

\definecolor{darkred}{rgb}{0.75,0,0}

\newcommand{\cmark}{\ding{51}}
\newcommand{\xmark}{\ding{55}}

\newcommand{\summarybox}[4]{
    \begin{center}
        \begin{tikzpicture}
            \node[anchor=text,text width=\columnwidth-1.2cm, draw, rounded corners, line width=1pt, fill=#3, inner sep=5mm] (big) {\\#4};
            \node[draw, rounded corners, line width=.5pt, fill=#2, anchor=west, xshift=5mm] (small) at (big.north west) {#1};
        \end{tikzpicture}
    \end{center}
}

\definecolor{sh_comment}{rgb}{0.12, 0.38, 0.18}
\definecolor{sh_keyword}{rgb}{0.37, 0.08, 0.25}
\definecolor{sh_string}{rgb}{0.06, 0.10, 0.98}
\definecolor{KWColor}{rgb}{0.5,0,0.67}
\definecolor{CommentColor}{rgb}{0.15,0.5,0.15}
\definecolor{lightgrey}{rgb}{0.8,0.8,0.8}

\lstdefinelanguage{swift}{
    morekeywords={
            open,catch,@escaping,nil,throws,func,if,then,else,for,in,while,do,switch,case,default,where,break,continue,fallthrough,return,
            typealias,struct,class,enum,protocol,var,func,let,get,set,willSet,didSet,inout,init,deinit,extension,
            subscript,prefix,operator,infix,postfix,precedence,associativity,left,right,none,convenience,dynamic,
            final,lazy,mutating,nonmutating,optional,override,required,static,unowned,safe,weak,internal,
            private,public,is,as,self,unsafe,dynamicType,true,false,nil,Type,Protocol,
        },
    morecomment=[l]{//},morecomment=[s]{/*}{*/},morestring=[b]",}

\lstdefinestyle{Eclipse}{
    xleftmargin=0pt,
    language=swift,
    basicstyle=\footnotesize\ttfamily,
    stringstyle=\color{sh_string},
    keywordstyle=\color{sh_keyword}\bfseries,
    lineskip=-0.0em,
    commentstyle=\color{sh_comment}\itshape,
    escapeinside={/*@}{@*/},
    numbersep=5pt,
    captionpos=b,
    xleftmargin=0.4cm, xrightmargin=0.5cm,
    morekeywords={invokestatic,invokeinterface,invokevirtual,invokespecial},
}

\newcommand{\rqone}{Which use cases do the current SBOM tools support?}
\newcommand{\rqtwo}{How is the health of current OSS SBOM tools distributed?}
\newcommand{\rqthree}{What are the areas for improvement in the OSS SBOM tools?}
\newcommand{\rqfour}{How is the health of GitHub projects that use SBOM tools distributed?}
\newcommand{\spdx}{SPDX\xspace}
\newcommand{\cdx}{CycloneDX\xspace}
\newcommand{\collectdate}{January 27th, 2024}
 
\begin{document}

\title{The State of SBOM Tools: A Comparative Analysis of \spdx and \cdx}
\titlerunning{The State of SBOM Tools}

\author{
	Zhimin Zhao \and
	Abdul Ali Bangash \and
	Tongxu Ge \and
	Arshdeep Singh \and
	Zitao Wang \and
	Bram Adams
}

\institute{
	Zhimin Zhao \at
	Software Analysis and Intelligence Lab (SAIL)\\
	Queen's University, Kingston, ON, Canada \\
	\email{z.zhao@queensu.ca}
	\and
	Abdul Ali Bangash \at
	Lahore University of Management Sciences\\
	Lahore, Pakistan \\
	\email{abdulali@lums.edu.pk}
	\and
	Tongxu Ge \at
	Software Analysis and Intelligence Lab (SAIL)\\
	Queen's University, Kingston, ON, Canada \\
	\email{tongxu.ge@gmail.com}
	\and
	Arshdeep Singh \at
	Indian Institute of Technology Ropar (IIT-Ropar)\\
	Ropar, Punjab, India \\
	\email{2020csb1074@iitrpr.ac.in}
	\and
	Zitao Wang \at
	Software Architecture Group (SWAG)\\
	University of Waterloo, Waterloo, ON, Canada \\
	\email{z254wang@uwaterloo.ca}
	\and
	Bram Adams \at
	Lab on Maintenance, Construction and Intelligence of Software (MCIS)\\
	Queen's University, Kingston, ON, Canada \\
	\email{bram.adams@queensu.ca}
}

\date{Received: date / Accepted: date}

\maketitle

\begin{abstract}

    A Software Bill of Materials (SBOM) provides transparency and accountability for a software release by documenting the metadata of its components and their dependencies (e.g., version numbers), and as such is an essential component of modern software development. However, the adoption and utility of SBOMs depend heavily on the tools that generate, analyze, and manage them. Software Package Data Exchange (\spdx) and \cdx are the two leading SBOM formats, yet the tools supporting them may differ in maturity, features, and community engagement. Since understanding the strengths and weaknesses of these tools is vital to improving SBOM practices and guiding future development, we first conduct a quantitative comparison of use cases for the $108$ open-source software (OSS) and $62$ proprietary SBOM tools publicly advertised for \spdx and \cdx, identifying areas where tools of each format may require enhancements in their features. Second, we compare the community-health metrics of each format's OSS tool repositories on GitHub, as collected from the official tool centers and contributing forks ($145$ \cdx and $456$ \spdx repository-format observations, covering $573$ distinct repositories), to evaluate their relative robustness and maturity. Additionally, to provide concrete insights about the challenges and potential development areas of each format, the study quantitatively compares $26{,}853$ format-specific observations of issue reports created since 2018 in these \cdx and \spdx tool repositories and obtains category assignments for $26{,}772$ of them with a large language model (LLM)-assisted approach calibrated against manually checked annotations. Finally, we investigate the characteristics of the top $250$ OSS projects that use \cdx tools and compare their community-health metrics with the top $250$ OSS projects that use \spdx tools. Our findings, which describe the tools listed in the official tool centers and the GitHub repositories identifiable within our data collection window, reveal distinct characteristics for each format: while projects that use \cdx tools often demonstrate higher developer engagement and certain community-health metrics, \spdx has more OSS tool repositories. This research contributes insights for software developers, OSS contributors, and SBOM practitioners regarding the complementary strengths of the \spdx and \cdx tools and identifies opportunities for mutual enhancement across both formats.
    
\end{abstract}

\section{Introduction}
\label{sec:intro}

Modern software applications rely heavily on third-party dependencies, forming the foundation of their software supply chain~\citep{cox2019surviving}. 
However, the increasing number of dependencies poses significant challenges in terms of security and reliability~\citep{gkortzis2021software}. 
Malicious actors are known to exploit vulnerabilities in less secure dependencies, ultimately tainting the supply chain and facilitating the breach of a high-value application via a compromised dependency~\citep{ladisa2022taxonomy}.
To address such vulnerabilities, it is crucial to have up-to-date information about all third-party dependencies in the supply chain of an application. 

In the past, the absence of proper formats for managing such information has resulted in serious consequences. 
Notably, in 2023, OpenAI temporarily took ChatGPT offline after a bug in redis-py, an open-source Redis client library, exposed some users' data to other users~\citep{openai2023outage}. 
Similarly, in the 2020 SolarWinds attack, a compromised update of the Orion platform may have been installed by fewer than 18,000 of its customers~\citep{solarwinds2020form8k}, highlighting the need for robust supply chain security. 
Moreover, the Log4Shell vulnerability in 2021 exploited the widely used Log4j library, necessitating prompt actions to protect systems and data~\citep{dynatraceWhatLog4Shell}. 

To mitigate the risk of such software supply chain attacks, in 2021, the US government issued an executive order requiring Software Bills of Materials (SBOMs) for software sold to the federal government~\citep{eo14028}, an endeavor followed later by regulatory proposals in, among others, the European Union and the United Kingdom~\citep{forrester,cra2024}. The concept of SBOM came from the notion of \emph{bill of materials} in other engineering disciplines~\citep{boms}.
Essentially, an SBOM is a precise specification of all ``ingredients'' of a given (software) product, ranging from the names and revision numbers of all source code and other files shipped with a given software release, to the names and version numbers of all third-party dependencies the product relies on, or even the licenses of those dependencies. 

To understand the readiness of SBOMs in the software development community, in 2021, the Linux Foundation surveyed 412 organizations from around the world (reported in 2022)~\citep{linuxReadiness}. 
Their survey showed that more than 80\% of worldwide organizations are aware of the US executive order and 76\% of organizations are considering changes to their cybersecurity strategies as a consequence of this executive order.
Overall, 47\% of organizations want scalable vulnerability reporting and 45\% see SBOMs as a key method to secure the software supply chain. 
They also found that 52\% of organizations are addressing SBOMs in at least a few areas of their business. 23\% are even addressing them across nearly all areas of their business or have standard practices that include the use of SBOMs. With the ever increasing understanding of SBOMs in general public, customers have started to require SBOMs from the developers of software products they use~\citep{linuxReadiness}. 

The specific information included in an SBOM relies on the SBOM format being used. The U.S. National Telecommunications and Information Administration (NTIA)'s SBOM specification outlines which data fields must be included in an SBOM file, how SBOMs should support automation, and which practices and processes should be employed when creating, distributing, and using SBOMs~\citep{NTIA2021}.
According to NTIA~\citep{NTIA2021}, three existing formats meet their requirements: \cdx~\cite{CycloneDXSpecs}, Software Package Data Exchange (\spdx)~\citep{spdx_about}, and Software Identification (SWID)~\citep{waltermire2016improving}.

While today's leading SBOM formats provide structured guidelines for the data representation of software components~\citep{bi2023way}, the formats by themselves have no value without the availability of appropriate tools that support and enforce its technical implications. For instance, one must have SBOM tools to generate, edit, visualize, and manage an SBOM of a software project. The build-supporting SBOM tools enable automatic extraction of key information about a software product release (such as the source code files and dependencies) during its build process. Similarly, inspection-supporting SBOM tools can be used to inspect compliance between the SBOM format and the release contents of a product, ensuring that all the components meet licensing and security requirements. This, in turn, requires more SBOM tools and libraries that can parse SBOMs across various programming languages.

However, in the pursuit of employing the right SBOM tools for their use cases, software development teams encounter various hurdles, such as limited standardization across formats and tool maturity, as neither of the two leading SBOM formats (\spdx and \cdx) fully meet current market needs for extensibility across software projects~\citep{xia2023empirical}. \cite{xia2023empirical} in their comprehensive survey on SBOM practitioners (17 interviewees and 65 survey respondents) found that current tools need to be more reliable, user-friendly, and compliant with the official formats. 
They found that practitioners currently adopt SBOM tools in a limited capacity due to a lack of understanding about these tools and their integration (both within an organization's own infrastructure, and amongst each other), including their technical aspects, use cases, resource costs, drawbacks, and benefits. 
They also found that the effectiveness and sustainability of SBOM tools are crucial for ensuring the security and compliance of software supply chains. Lastly, there is generally a limited availability, usability, and integration support for SBOM tools~\citep{xia2023empirical}.

Similar to what the Linux Foundation's Community Health Analytics in Open Source Software (CHAOSS)~\citep{chaoss} initiative is doing for open-source software (OSS) projects in general, understanding the current SBOM tools, such as their supported formats, use cases, their health, issue reports, and demands, would help SBOM practitioners select the right SBOM tools for their use case, while it would enable SBOM tool developers to improve their tools. Hence, in this paper, we empirically study the tools supporting the two leading SBOM formats, \spdx and \cdx, which prior work identifies as the most adopted formats~\citep{xia2023empirical} and the leading formats among SBOM tools~\citep{mirakhorli2024landscape}. Throughout, ``\spdx tools'' and ``\cdx tools'' refer to the tools listed in the official \spdx and \cdx tool centers and, for repository-level analyses, to their GitHub-hosted repositories and contributing forks identifiable within our data collection window. Our findings do not cover tools outside these sources. We investigate the following research questions (RQs):

\begin{itemize}
    \item RQ1: \rqone
    \item RQ2: \rqtwo
    \item RQ3: \rqthree
    \item RQ4: \rqfour
\end{itemize}

In \textbf{RQ1}, we manually investigate which SBOM use cases the current OSS and proprietary SBOM tools support. 
We find that proprietary tools are richer than OSS SBOM tools when it comes to the number of use cases they can support.
Furthermore, we identify the use cases with which OSS tools are negatively associated.
Finally, we highlight the different strengths and weaknesses of each of the formats' (\spdx and \cdx) tools.

In \textbf{RQ2}, to assess the overall health of OSS \spdx and \cdx GitHub tools, we evaluate them using CHAOSS community-health metrics. Our findings reveal notable differences between the two formats' tools, with \cdx tools generally exhibiting higher contributor activity and more frequent commits. While these observations reflect the state of the tools at the time of our analysis, they also point to a broader insight: the vitality and maintenance of supporting tools can vary substantially between formats. Therefore, developers should consider not only the SBOM format itself but also the maturity and health of its supporting tools when selecting a format for their use case.

In \textbf{RQ3}, to understand the potential areas of improvement in the tools of each format, we extract and investigate the GitHub issue reports of the \spdx and \cdx SBOM tools.
We derive a 14-category taxonomy from maintainer-assigned tags and use a large language model (LLM)-assisted approach, calibrated using a sample of 379 issue reports, to assign categories to 26,772 of the 26,853 format-specific issue-report observations created since 2018. We then compare category prevalence and resolution time between the formats.
The tools supporting \cdx have more closed issue reports and a shorter mean resolution time than those supporting \spdx, although the overall resolution-time effect size is negligible. 

In \textbf{RQ4}, to understand the types of projects that adopt specific SBOM tools, we analyze OSS GitHub projects that have integrated SBOM tools within their continuous integration (CI) pipelines to automate the generation of SBOM files during the build process. We observe that projects with a higher number of commits tend to use SBOM tools that support the \emph{View}, \emph{Diff}, and \emph{Import} use cases, whereas projects in the second commit quartile more often use tools that support \emph{Analyze}, \emph{Edit}, and \emph{Merge}. Additionally, we find that more active projects, those with higher numbers of stars, watchers, releases, pull requests (PRs), and contributors, are more likely to use \cdx tools.

Rather than focusing solely on SBOM formats, our empirical study highlights the critical role of the supporting tools. By offering an overview of the tools of the two leading SBOM formats, \spdx and \cdx, we examine their suitability for different use cases and identify common challenges, thereby enabling practitioners to make informed decisions when selecting tools for SBOM management. Our findings reveal differences in the populations of projects adopting each format's tools, with variations in developer engagement and integration practices. These observations underscore that the choice of an SBOM format should be considered alongside the maturity and alignment of its tools with the needs of a given project.
The data and analysis artifacts supporting this study are publicly available in our replication package~\citep{replication}.
 \section{Background and Related Work}
\label{sec:backgr-relat-work}

In this section, we explain what an SBOM is, what is an SBOM specification, and which are today's leading SBOM formats.
We then explain how these concepts are related to our work and what kind of relevant studies have been performed in the literature.

\subsection{What is an SBOM format?} 
The concept of a bill of materials (BOM) originated from the manufacturing industry's practice of creating a document that records and helps to track and manage components used in the production of physical goods~\citep{boms}.
Drawing inspiration from this traditional notion of BOM, the notion of a software bill of materials (SBOM) was defined for software engineering as early as 2007 by~\cite{van2007component}.
An SBOM is a structured document that records the detailed information about the components and dependencies comprising a software application or system~\citep{muiri2019framing}. 
It keeps the names, versions, and sources of third-party libraries, frameworks, modules, and other software elements used in the development process. 
SBOM serves as a comprehensive record of the software supply chain, enabling organizations to
uniquely and unambiguously identify components of an application and their relationships, to understand and manage the security, licensing, vulnerability, and compliance aspects of their software assets. In practice, SBOMs can be created at different stages of the software lifecycle, leading to distinct SBOM types, such as Design, Source, Build, Analyzed, Deployed, and Runtime SBOMs, each with its own benefits and limitations depending on how and when the SBOM is generated\footnote{\url{https://www.cisa.gov/sites/default/files/2023-04/sbom-types-document-508c.pdf}}.

An SBOM specification, often developed by a regulation authority, outlines the rules and information that should be included in an SBOM to ensure accurate identification and tracking of software components and dependencies. The SBOM specification typically includes details on which software assets should be described, organized, and documented within the bill of materials. 
An example of an SBOM specification is the U.S. NTIA's SBOM specification, which outlines the data fields that must be included in an SBOM file, how SBOMs should support automation, and which practices and processes should be employed when creating, distributing, and using SBOMs~\citep{NTIA2021}.

The NTIA recognizes three main SBOM formats that follow their SBOM specification: \spdx, \cdx, and SWID~\citep{NTIA2021}. 
Each format has its unique focus and strengths, catering to different stages of the software development lifecycle and varying user needs. The \spdx format, initiated by the Linux Foundation and currently developed in collaboration with industry experts, organizations, and OSS enthusiasts through the community-driven \spdx Project~\citep{spdx_ref}, is used, for example, by Microsoft for all the SBOMs it generates~\citep{diglio2021spdx}. On the other hand, the \cdx format, developed within OWASP\footnote{\url{https://owasp.org}} as a security-focused bill of materials standard, aims at application security and supply chain component analysis~\citep{cyclonedx_ref}.
Similarly, SWID (Software Identification tags) is another format to identify and track software installations, facilitating software asset management and cybersecurity efforts~\citep{swid_ref}.
In this study, we focus on the two leading SBOM formats: \spdx and \cdx~\citep{ravi}. Notably, \spdx (established 2010~\citep{spdx_ref}) is more mature than \cdx (introduced 2017~\citep{cyclonedxspec}), which may influence the nature and volume of issue reports reported for tools supporting each format.

\subsection{Related Work}
\label{sec:related-work}

\cite{9174365}, based on their industrial experience, discuss 11 crucial elements that an SBOM format should have and also highlight nine possible use case scenarios of an SBOM.
In our work, we identify the use case scenarios addressed by tools listed in the official \spdx and \cdx tool centers. Our findings therefore characterize these listings and the GitHub-hosted repositories and contributing forks identifiable within our data collection window, rather than the entire SBOM tool ecosystem. Instead of using Martin~\textit{et al.}'s taxonomy, we use the NTIA tool taxonomy, which is a collective effort of practitioners and tool developers across multiple SBOM formats.

\cite{xia2023trust} addressed the limitation of secure SBOM sharing among software vendors in their work. They identified the SBOM sharing process of the current SBOM tools that support the \cdx format and proposed a blockchain-based solution with verifiable credentials for secure and flexible disclosure. Their approach enhances the security, usability, and privacy of the SBOM sharing process between the software vendors. 

\cite{xia2023empirical} explored the importance of SBOMs in strengthening the security of the software supply chain. Through interviews with 17 practitioners and surveys of 65 respondents from 15 different countries, they identified the current challenges SBOM practitioners face. In addition to general problems related to SBOM, they found limited adoption of SBOM in the software community due to the community's lack of understanding of the technical aspects of SBOMs and the use cases that SBOM tools offer. They highlighted that current SBOM tools suffer from issues in reliability, user-friendliness, and compliance with official formats. Furthermore, they emphasized the critical need for a robust SBOM tool ecosystem. We complement their practitioner-level investigation by examining supported use cases and repository issue reports at the level of the tools and their repositories.

\cite{bi2023way} perform a comprehensive survey on 4,786 GitHub discussions of projects that use SBOMs. Their work highlights key topics, challenges, and solutions intrinsic to the effective utilization of SBOMs. However, in this paper, we focus on the topics and challenges of the adoption of SBOM tools instead of SBOM format usages in general.

\cite{nocera2023software}'s work is closely related to our RQ4, as they examine the use of SBOM tools in 186 GitHub projects. However, their analysis is limited by the fact that SBOM files are rarely included directly in repositories and are typically only bundled in generated releases, resulting in a relatively small number of data points. In contrast, we identify projects that actively use SBOM tools in their CI pipeline, allowing for broader and more systematic coverage. Moreover, while their study focuses on the frequency of SBOM file updates, it does not explore the characteristics of projects that adopt these tools. They also do not distinguish between the \spdx and \cdx formats in their analysis, and their findings remain largely demographic in nature.

\cite{mirakhorli2024landscape}'s work is closely related to our work's RQ1, in which they evaluate 84 OSS and proprietary SBOM tools based on their support for SBOM formats, data completeness, tool characteristics (e.g., license and cost), and additional features such as vulnerability and license management. To identify their emerging use cases, they execute 5 SBOM tools and demonstrate their interoperability problems. Our study, in contrast, identifies the use cases for 170 distinct SBOM tools (from 187 tool-center entries) available at the time of analysis, employing the NTIA tool taxonomy.
Furthermore, instead of carrying out a qualitative investigation of five tools, we analyze issue reports from 261 repository-format observations since 2018 (237 distinct GitHub repositories) to study the most prevalent problems within them.
Lastly, our study extends beyond the investigation of use cases and problems of SBOM tools, as it further looks into the health factors of the projects that use these tools.

The CHAOSS metrics used in \textbf{RQ2} and \textbf{RQ4} to assess the overall health of OSS software projects have also been used in previous research to study ecosystem health~\citep{mens2017towards} and contributor onboarding in OSS ecosystems~\citep{foundjem2021onboarding}. The onboarding study applied observational and empirical methods to investigate onboarding strategies, challenges, and benefits, as well as their effects on diversity, productivity, and code quality. 
 \section{Overview of the Study}

\begin{figure}[t]
    \centerline{\includegraphics[width=\linewidth]{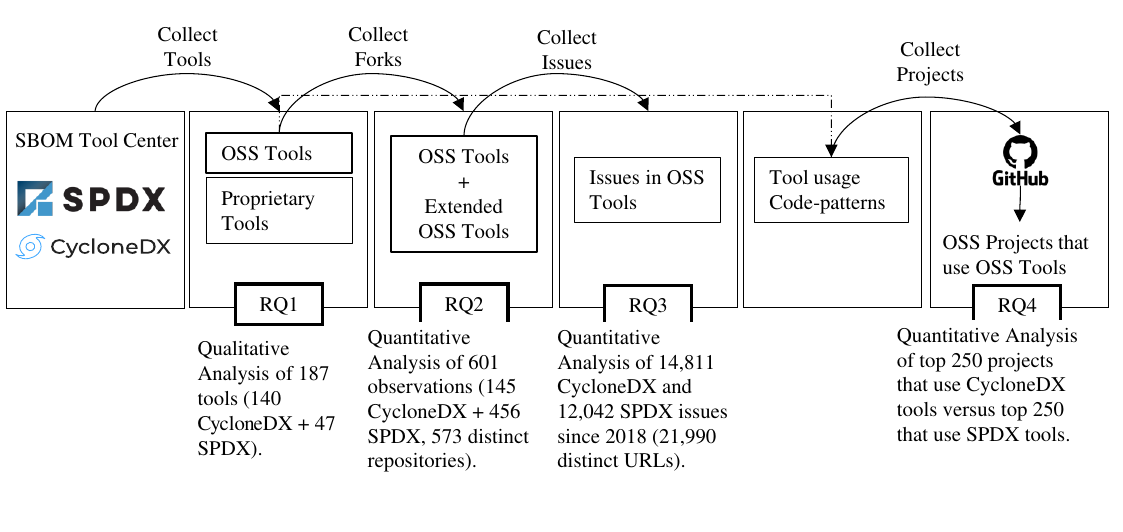}}
    \caption{Overview of our methodology and dataset.}
    \label{fig:method}
\end{figure}

Figure~\ref{fig:method} provides a general overview of our methodology to address the RQs listed in the introduction.
Our dataset consists of the tools listed in the official tool centers (snapshot of \collectdate), their GitHub repositories, and the contributing forks identifiable on GitHub within the data collection window.
First, we collect all the SBOM tools available at the time of our research from the tool center website of both studied formats (\spdx~\footnote{\url{https://spdx.dev/spdx-tools/} (at the time of collection, \url{https://spdx.dev/tools-community/})} and \cdx~\footnote{\url{https://cyclonedx.org/tool-center/}}), we then qualitatively analyze these tools to identify which use cases they support and how they are different from one another (RQ1).
Second, we extend the initially collected OSS tools dataset with OSS tools (including forks) available on GitHub to compare the tools' project health between both formats (\spdx and \cdx) quantitatively (RQ2).
Third, we extract the GitHub issue reports of all the collected OSS tools and quantitatively analyze the most prevalent type of issue reports (RQ3).
Finally, we qualitatively curate a list of search patterns from the previously collected list of OSS tools to identify projects on GitHub that use these tools. After collection, we study the characteristics of these projects to understand which types of projects prefer adopting specific types of tools across both formats (RQ4). We collected the repository and project metrics of RQ2 and RQ4. Note that we normalize these community-health metrics by repository age (in days) to account for variations in project maturity and ensure fair comparisons between repositories of different lifespans. RQ3 instead restricts both formats' repositories to the same window since 2018 and compares raw per-repository issue report counts.

\subsection{Collecting the SBOM tools }
For RQ1, we extracted all the tools advertised on the official tool center websites of \cdx~\citep{CycloneDXTools} and \spdx~\citep{spdx_tools_opesource}.
We found 219 tools on the \cdx tool center website and 47 tools on the \spdx tool center website. 
On closer inspection, we found some discrepancies in the information provided on the websites of both the tool centers, i.e., we found that not all the tools mentioned as OSS by their developers are necessarily OSS. Some of the tools have most of their functionalities private and/or commercial. 
Secondly, we found that 78 tools listed across the \cdx and \spdx tool centers are dual-format, either explicitly supporting both formats or having use cases listed in the \spdx tool center that do not fully reflect the broader range of functionalities they provide.
We also identified certain \spdx tools whose use cases, as listed on the \spdx tool center, may not reflect the complete range of use cases the tool supports.

\subsection{Manual Analysis to identify tool use cases }

To determine the complete range of the use cases of these tools, and to determine which ones are truly OSS, we manually analyze them as follows. With a 95\% confidence interval and an error margin of 5\%~\citep{cochran1977sampling}, we randomly selected 140 tools from the 219 tools that were available on \cdx tool center, and selected all 47 tools that were available on the \spdx tool center. This step leaves us with 187 tools to analyze in total.
Two of the authors manually investigate the tools to categorize them as OSS/proprietary and to determine whether they support a single format or both (\spdx and \cdx). 
In addition, to highlight the use cases that a tool supports, the same two authors, using the closed-card sorting~\citep{zimmermann2016card} technique, categorized these tools according to the official NTIA tool taxonomy~\citep{NTIATaxonomy}.
This tool taxonomy, which we discuss below, was created by NTIA after surveying SBOM producers, consumers, and tool developers to provide a variety of SBOM management use cases~\citep{sbommeeting}.
We chose the NTIA tool taxonomy as it is the result of a collective effort by multiple SBOM format practitioners and tool developers~\citep{sbommeeting}, and has been adopted as the authoritative framework for SBOM tool classification by the broader SBOM community~\citep{mirakhorli2024landscape,xia2023empirical}.
The NTIA tool taxonomy comprises the following categories of tools:

\begin{itemize}[leftmargin=*]
\item The \textit{Produce} category includes tools that automatically create SBOMs during software artifact creation, analyze source or binary files to generate SBOMs, and facilitate manual data entry.
\begin{itemize}
\item \textbf{Build} - Tools, such as Valaa Stack\footnote{\url{https://github.com/valaatech/kernel/blob/8e24e9e731bb67d0da82fbd56b6a8e32f22c178f/CHANGELOG.md?plain=1\#L841}}, that automatically generate an SBOM as part of the software build process, containing information about the build.
\item \textbf{Analyze} - Tools, such as SecObserve\footnote{\url{https://github.com/MaibornWolff/SecObserve/blob/27db653939e6209f8b6a24d3a45a76c08f358086/docs/usage/supported_scanners.md}}, that analyze source or binary files to generate an SBOM by inspecting artifacts and associated sources. 
\item \textbf{Edit} - Tools, such as CycloneDX PHP Library\footnote{\url{https://github.com/CycloneDX/cyclonedx-php-library}}, that assist in manually entering or editing SBOM data.
\end{itemize}
\item The \textit{Consume} category includes tools that allow users to view SBOM contents in a human-readable form, compare multiple SBOMs to identify differences, and import SBOMs for further processing and analysis.
    \begin{itemize}
        \item \textbf{View} - Tools, such as Tidelift\footnote{\url{https://support.tidelift.com/hc/en-us/articles/4406286154004-About-projects-and-bill-of-materials\#obtaining-a-bill-of-materials-0-1}}, that display SBOM contents in a human-readable form, supporting decision-making and business processes.
        \item \textbf{Diff} - Tools, such as SBOMTrend\footnote{\url{https://github.com/anthonyharrison/sbomtrend}}, that compare multiple SBOMs and highlight differences, such as between different versions of a software product.
        \item \textbf{Import} - Tools, such as NetRise Turbine\footnote{\url{https://www.netrise.io/platform/software-bill-materials-management}}, that enable the discovery, retrieval, and import of SBOMs into systems for further processing and analysis.
    \end{itemize}
\item The \textit{Transform} category includes tools that translate SBOMs between file types, merge multiple SBOM sources for analysis and audit, and support integration with other tools through APIs and object models.
    \begin{itemize}
        \item \textbf{Translate} - Tools, such as dbom\footnote{\url{https://github.com/HaRo87/mdbom}}, that convert an SBOM from one file type to another while maintaining the same information.
        \item \textbf{Merge} - Tools, such as SBOM Assembler\footnote{\url{https://github.com/interlynk-io/sbomasm}}, that combine multiple SBOM sources and other data for comprehensive analysis and auditing. 
        \item \textbf{Tool Support} - Tools, such as Enso\footnote{\url{https://knowledge.enso.security/connectors-and-integrations/connectors-and-marketplace}}, that facilitate SBOM use in other applications through APIs, object models, libraries, and other reference sources.
    \end{itemize}
\end{itemize}

To avoid bias in manual classification, two authors initially categorized a random sample of 29 out of 187 tools based on their use cases. During this process, the authors documented the different sources required for categorization. Each rater independently assigned a binary (yes/no) label to each use-case dimension per tool, treating the nine dimensions as separate binary decisions. In our first round of manual classification, the inter-rater reliability across those $9 \times 29$ ratings yields a Krippendorff's alpha~\citep{krippendorff2019content} of $0.734$, indicating ``substantial agreement'' between raters on the scale of Landis and Koch~\citep{landis1977measurement}. Following the initial round, the authors discussed their categorization approaches, resolved conflicts through negotiation, and established a unified methodology for analyzing the remaining 158 samples. In the final round of this manual analysis, the inter-rater reliability agreement yields a Krippendorff's alpha of $0.872$, indicating ``almost perfect agreement'' on the same scale. The remaining conflicts in the classification of tools were resolved through negotiated agreement~\citep{campbell2013coding}. In Appendix~\tabref{tab:tool-classification}, we present the results of the 170 distinct tools in our sample, of which 62 are proprietary and 108 are OSS.

The manual analysis involved examining tool descriptions on the \cdx or \spdx tool center websites, supplemented by additional exploration based on tool type. For OSS tools (available on GitHub), we reviewed repository README files. When READMEs provided insufficient information, we searched repositories using keywords ``sbom,'' ``spdx,'' and ``cycloneDx,'' then examined relevant Markdown or text files. If such files were unavailable, we analyzed code usage of ``spdx'' and ``cycloneDx'' keywords, which typically revealed whether tools could generate, consume, or transform SBOMs and indicated format support capabilities.

For proprietary tools, we search for the keywords ``sbom'', ``spdx'', and ``cycloneDx'' initially on the landing web-page of the tool, which would sometimes provide us with information on the use cases and format supported by the tool. If not, we search the same keywords along with the tool name on the Google search engine. This search would yield information from a variety of sources, ranging from LinkedIn or Medium blogposts written by the tool developers to the tool's documentation and demo tutorials. In case we find a tool demo or documentation, we read through it to understand its use cases, while in the other case, when there is only a blog post available, we read through the blog post to understand what the tool does.

Both formats' tool centers support tags along with tool names as well. While these helped us identify the primary use case of the tool, on closer inspection, we found that the tags do not provide a sufficiently complete picture of which use cases the tool supports, and hence we did not consider tags in the remainder of our analysis.
Both tool centers also had a number of duplicate tools with different naming conventions. Therefore, we mapped the official URLs of the tools and identified 18 duplicates, with one exception where one URL carried two different tools.

\subsection{Extending the collection of OSS SBOM tools }
\label{sec:method:tools-extension}
For RQ2, we expanded on the collection of current OSS tools available in the \spdx and \cdx tool center by extracting their forks from the GitHub repositories.
One of the authors manually analyzed the forked repositories to identify contributing forks\footnote{\url{https://chaoss.community/kb/metric-technical-fork}} and filtered out forks that did not expand over the parent tool, i.e., without further commits after the fork (non-contributing forks). They also filtered out the forks without a description in the README file, since those do not allow us to know if it is a tool repository or not.
Contributing forks are excluded from RQ1 to avoid duplicating parent tools in the use-case analysis, but included in RQ2 and RQ3 to capture the full scope of active development for each format.

As a result, collectively from the tool centers and the contributing forks, we found $145$ \cdx and $456$ \spdx repository-format observations. Since 28 repositories support both formats and therefore appear in both groups, these $601$ observations cover $573$ distinct GitHub repositories.
Solely based on the tool-center data, \cdx had more OSS tools available. However, after including the contributing forks of the tools for each format, the total number of OSS tools for \spdx increased.
A larger set of tools could suggest a wider variety of functionalities and potential benefits for SBOM practitioners.

We also scanned the 456 \spdx tool repositories for \spdx 3.0 keywords to assess the uptake of that version (Section~\ref{sec:threats}).

\subsection{Extracting the issue reports and their tags of the OSS SBOM tools }
\label{sec:method:tag-grouping}
For RQ3, we queried the GitHub repositories behind the 145 \cdx and 456 \spdx repository-format observations to collect their issue reports.
Since not all repositories contain issue reports, we collected 17,795 \cdx and 19,194 \spdx issue reports. Since some tool repositories appear more than once in our tool lists (Section~\ref{sec:method:tools-extension}), the same issue report appeared up to three times within a format, so we kept each issue report once per format. This leaves 15,512 distinct issue reports in 90 \cdx tool repositories and 13,839 in 190 \spdx tool repositories.
Since \cdx's first prototype was released in May 2017, we filter out the issue reports, both from \cdx and \spdx GitHub tool repositories, that were created before 2018. This cutoff ensures that both formats' tools are measured over a comparable active development period (i.e., when both formats were actively maintained and their histories overlap), excluding \spdx's pre-2018 history as well as \cdx's brief pre-2018 period. Within this window, we compare raw per-repository issue report counts without age normalization, so repositories that are older within this period (on average the \spdx ones) have had more time to accumulate issue reports.
After this filtration step, we were left with 14,811 closed and open issue reports in 90 \cdx tool repositories and 12,042 closed and open issue reports in 171 \spdx tool repositories. Since our comparison is between formats, an issue report from a repository that supports both formats contributes once to each format. The resulting 26,853 format-specific observations correspond to 21,990 distinct issue-report URLs.

\textbf{Step 1: Deriving the issue category taxonomy from tagged issue reports.}
To build a category taxonomy, we first examine issue reports that carry tags assigned by tool maintainers on GitHub, yielding 13,352 tags associated with 8,257 ($68.57\%$) issue reports of 91 \spdx tool repositories (361 unique tags) and 19,682 tags associated with 11,958 ($80.74\%$) issue reports of 65 \cdx tool repositories (360 unique tags).
We sorted tags by frequency and analyzed their distribution. Within the fourth quartile (top 25\% most frequent tags), we found that for \cdx the 90 most common tags appear in 24 to 4,933 issue reports, while for \spdx the 91 most common tags appear in 21 to 2,703 issue reports. The first author performed the following curation on the tag-frequency list as it stood before duplicate removal (all curated tags remain present after removal, and 83 of the 95 distinct tag--format pairs remain in the top quartile). The first author manually investigated the most common \cdx and \spdx tags to exclude those that do not convey anything about the meaning of an issue report, such as ``P1'', ``minor'', ``cannot reproduce'', ``super'', ``easy'', and ``not an issue'', etc. Furthermore, tags with the same meaning were merged, such as ``technical debt'' and ``technical-debt'', ``license scan'' and ``license'', and ``new feature'' and ``nice to have'' etc. This process leaves us with the 49 most prevalent unique tags in \cdx tools and 48 most prevalent unique tags in \spdx tools. The complete list of tags and their categories is available on our replication package.
After merging the tags, we manually abstract similar tags into broader categories. For this process, the first author sorts the tags and groups the ones that have similar purposes, such as ``defect'', ``technical issue'' and ``bug'', categorized as ``Bug Fixes and Defects'', docs and documentation categorized as ``Documentation'', and ``enhancement'', ``must have'', ``nice to have'' and ``feature'', categorized as ``Feature Development and Enhancement''. The last author then, having the most experience, re-analyzed the categories of the tags and raised 10 concerns about the grouping scheme. These concerns were related to: the possible duplication of tags, such as \cdx having duplicate tags named ``stale'', mis-judgement of categories e.g., ``docs'' falling into ``CI/CD'' category instead of ``Documentation'', and better naming convention, e.g., ``Bug Fix'' should be ``Bug Fixes and Defects''. Through a negotiated discussion, the authors refined the 97 curated tag entries (\cdx: $49$, \spdx: $48$, or 95 distinct tag--format pairs since some tag names repeat within a format) into 14 distinct categories, as detailed in Section~\ref{sec:rq3-result}.

\textbf{Step 2: Calibrating the LLM prompt using manual annotations.}
With the 14-category taxonomy established, we constructed a prompt for \texttt{mistralai/devstral-small-2-2}~\footnote{\url{https://huggingface.co/mistralai/devstral-small-2-2-24B-Instruct-2512}} (served via a local vLLM endpoint) to classify each issue report into one or more of the 14 categories based solely on its title and body text. The prompt contains the category definitions, classification rules, and three representative few-shot examples, and we ran the model at temperature 0 on the issue title and the first 1,600 plus the last 400 characters of the body. We selected a sample of 379 issue reports for prompt calibration, with the sample size determined using Cochran's criterion~\citep{cochran1977sampling} for a 95\% confidence level and a $\pm5\%$ margin of error. The first author iteratively refined the category descriptions and classification rules by comparing the model's outputs with the available manual multi-label annotations based on issue titles and bodies. After manual checking and iterative prompt refinement, agreement between the LLM classifications and the manually checked annotations reached Krippendorff's alpha~\citep{krippendorff2019content} of $0.731$ on the calibration sample. The replication package records annotation sources and distinguishes human annotations from model-generated annotations; only human annotations serve as manual overrides. Because this sample was used for prompt calibration, agreement on it is a development measure rather than an independent estimate of classifier performance. The 95\% confidence level and $\pm5\%$ margin describe the sample-size calculation, not a confidence guarantee for classifier accuracy or agreement. We therefore treat the LLM labels as approximate classifications for aggregate-level analysis.

\textbf{Step 3: Classifying issue reports across the dataset.}
\label{sec:rq3-llm-classify}
After prompt calibration, we applied the primary model to issue reports created since 2018, using their titles and bodies and the same category definitions and final prompt. This approach includes untagged issue reports (6,638, or 24.7\% of the 26,853 format-specific observations). The consolidated classification combines the primary model outputs with title/body annotations from \texttt{gpt-6-astra} via PackyAPI under the same taxonomy, with available human annotations applied as overrides. Categories mapped from maintainer-assigned tags are used for 23 observations. Annotation sources and model provenance are recorded in the replication package.

The final category analysis includes 26,772 format-specific observations (14,763 for \cdx and 12,009 for \spdx), or 99.7\% of the 26,853 observations. Category prevalence is the proportion of categorized issue reports within each format that carry a given category; an issue report with multiple categories contributes to each category but only once to that format's denominator. Resolution-time comparisons use closed issue reports. These assignments support the prevalence and resolution-time analyses in Section~\ref{sec:rq3-result}. Classifying titles and bodies using consistent category definitions reduces dependence on repository-specific tagging conventions, although the taxonomy is tag-derived and 23 assignments use tag-based fallbacks.

Applying the LLM classifier to tagged issue reports lets us cross-check its results against the tag-based classification, which uses the maintainers' tags (Section~\ref{sec:rq3-result}). Since the tag-based classification covers only the tagged issue reports ($68.57\%$ for \spdx) and depends on how each repository's maintainers apply tags, whereas the LLM-assisted approach provides category assignments for 26,772 format-specific observations using consistent category definitions and a prompt calibrated against available manual annotations in Step~2, we base the RQ3 comparisons on the LLM-assisted classification.

\subsection{Collecting OSS projects that use OSS SBOM tools }

For RQ4, an approach to collecting projects that use SBOM tools, previously employed by~\citep{xia2023empirical}, is to select GitHub projects that contain SBOM files in their repositories. However, our manual exploration revealed that, at the time of our analysis, projects rarely store SBOM files in version control. This is since SBOMs are typically generated for official releases, rather than for intermediate development snapshots (i.e., individual commits), and are often bundled within release artifacts rather than committed to the repository itself. Thus, we cannot easily find GitHub projects by searching for those that contain an SBOM file. 

Since our aim is to determine which SBOM tools are being used by GitHub projects in general, we searched for traces of command line invocations that would indicate these SBOM tools are being automatically executed within project CI pipelines. We focus on Build-use-case tools since SBOM generation is the primary use case exercised in CI pipelines. Other use cases such as Consume (viewing, importing) or Transform (translating, merging) are typically performed as post-build or ad hoc steps outside of automated pipelines. This focus is motivated empirically: CI systems are designed to generate build artifacts automatically, making Build the naturally leading use case.
To identify the search patterns for each SBOM tool, we analyze the codebase and usage guidelines in the README files of all OSS SBOM tools identified in RQ1 that explicitly support the Build use case from the NTIA tool taxonomy.

\label{sec:method:rq4-validation}
To validate this assumption and assess the precision of our collection approach, we drew a stratified random sample of 94 \texttt{tool\_mention\_link} entries (36 \spdx + 58 \cdx, proportional to format distribution), satisfying the Cochran criterion~\citep{cochran1977sampling} for a 95\% confidence interval with a $\pm10\%$ margin of error over the 2,426-project population. We fetched the raw file content from GitHub for each entry: 87 were successfully retrieved (7 returned HTTP 404, indicating deleted or now-private repositories). Of the 87 retrievable entries, 81 (93.1\% with a 95\% CI of 85.8\%--96.8\%) were confirmed as Build-use-case SBOM tool invocations through automated pattern matching (77) or manual inspection (4). The remaining 6 entries comprised 4 false positives, where the search pattern matched an unrelated file (e.g., a metadata JSON or documentation page), and 2 indirect references (files that import or describe the tool without directly invoking it). Crucially, \textbf{no Consume or Transform use-case invocations were found} in any of the 87 inspected files. The false positive rate is 4/87 = 4.6\% (95\% CI: 1.8\%--11.2\%).

We also note that the GitHub Search API returned matches across all file types: CI configuration files (28.5\% of all 3,458 matches), build scripts (Maven, Gradle, Make), source code files (.py, .java, .sh), and package manifests (JSON, YAML, TOML). We included all returned file types in our analysis, as build scripts, source files, and package manifests can all be invoked as part of a CI pipeline's execution. Of the validated entries, 93.1\% were confirmed as Build-use-case invocations, which supports the precision of our patterns. Since the patterns were derived from Build-oriented invocations, however, this does not show that Build is the leading use case outside our sampling frame. The full sample with per-entry evidence is included in our replication package (\texttt{RQ4-project/ci\_validation\_sample.csv}).
We identified the search patterns of 66 such OSS SBOM tools (11 for \spdx tools, 50 for \cdx tools, and 5 for both). 
All of these search patterns are available in our replication package, and we list a few in the Appendix~\tabref{tab:pattern_distribution}.
The search patterns may include dependency code of an SBOM tool, path to an SBOM tool library or a web service, or installation command of an SBOM tool.
To identify projects that use these SBOM tools, we then use the GitHub search API to identify projects that contain the search patterns of a tool in their code files or build configuration files. 

The search patterns for the projects that use \spdx tools or \cdx tools to create SBOMs returned $3{,}944$ raw matches. After removing the source repositories of the tools themselves, 3,458 tool-mention matches remained, representing $2{,}426$ distinct repositories: $307$ projects that use only \spdx tools, $1{,}087$ projects that use only \cdx tools, and $1{,}032$ projects that use tools for both formats. In our analysis, to compare \spdx and \cdx, we consider the $1{,}394$ projects that use only one of the two formats: $307$ use only \spdx tools and $1{,}087$ use only \cdx tools.
We then sort the projects by their number of GitHub stars and select the top 250 projects from each set. 
Selecting projects based on the number of GitHub stars ensures a focus on popular and well-supported projects, reflecting community interest and active development~\citep{borges2018s}.
Finally, we compare the project health of the top 250 most-starred projects that use \cdx tools with that of the top 250 most-starred projects that use \spdx tools.
 \section{Results}
\label{sec:eval-results}

\subsection{Use cases of the SBOM Tools (RQ1)}

In RQ1, we study the use cases of the SBOM tools of the \cdx and \spdx formats.
Investigating which use cases each format's tools support and what are the competing use cases among them would help us understand the scope and variety of the SBOM tools listed in the official tool centers.
A clear understanding of the strengths and weaknesses of each format's tools, in terms of the use cases they support, would allow SBOM practitioners to adopt the most suitable SBOM format and tools for their needs. Such an identification process would also highlight the capabilities and limitations of the available SBOM tools, which would help SBOM tool developers improve their tools.

\subsubsection{\spdx versus \cdx Tools.}

\textbf{Proprietary tools have a higher tendency to support more than one SBOM format than OSS tools.}
Across the 170 tools that we manually analyzed, 78 tools support both \spdx and \cdx, even if they are mentioned on the tool-center webpage of only one of the formats.
For comparison, we group the tools into two categories: single-format tools that support either \spdx or \cdx (92) and dual-format tools that support both formats (78).

\begin{figure}[t]
    \centerline{\includegraphics[width=\linewidth]{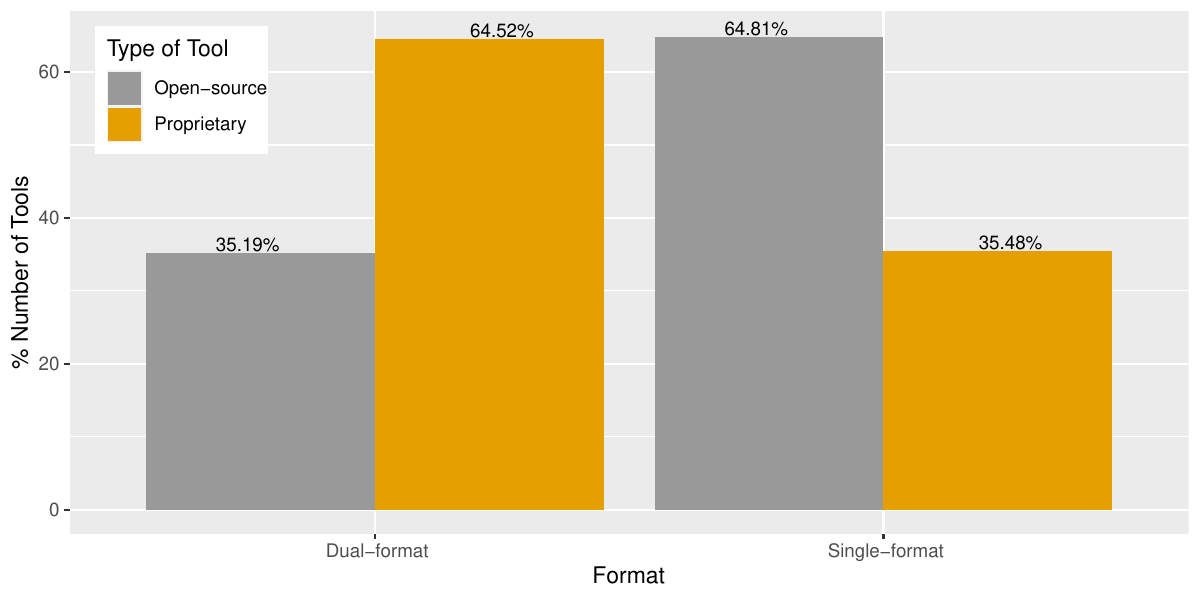}}
    \caption{Distribution of 170 manually analyzed SBOM tools across the SBOM formats that they support. Dual-format tools support both formats, while single-format tools support one format.}
    \label{fig:open_vs_prop_spdx_cdx_dual}
\end{figure}

Figure~\ref{fig:open_vs_prop_spdx_cdx_dual} shows that, among the tools that explicitly support either the \cdx or \spdx format, OSS implementations are the most prevalent. On the other hand, among the tools that support dual-format, proprietary is the most prevalent choice.
In other words, proprietary tools have a higher tendency towards supporting both formats (64.52\%) while OSS tools tend towards supporting a single format (64.81\%).

\summarybox{RQ1: OSS versus Proprietary in terms of format support.}{lightgray}{white}{Proprietary tools are more inclined towards supporting both formats (64.52\%), while OSS community tools are more inclined towards supporting a single format at a time (64.81\%).}

\begin{table}[t]
	\caption{Manual classification of 170 SBOM tools (92 single-format and 78 dual-format) according to the NTIA tool taxonomy.}
	\begin{center}
		\begin{tabular}{lllrr}
			\toprule
			\textbf{Category} & \textbf{Use case} & \textbf{Description} & \textbf{Dual} & \textbf{Single} \\
			\cmidrule{1-5}

			\multirow{3}{*}{Produce} 	& Build & SBOM creation & 74.36\% & 79.35\% \\
										& Analyze & Artifact analysis & 70.51\% & 57.61\% \\
										& Edit & Manual SBOM editing & 8.97\% & 6.52\% \\
										&	   & Total & 94.87\% & 92.39\% \\
										\cmidrule{1-5}
			\multirow{3}{*}{Consume} 	& View & Visualization of decision making & 67.95\% & 46.74\% \\
										& Diff & SBOM comparison & 10.26\% & 6.52\% \\
										& Import & SBOM retrieval & 43.59\% & 23.91\% \\
										&	   & Total & 79.49\% & 55.43\% \\
										\cmidrule{1-5}
			\multirow{3}{*}{Transform} 	& Translate & File type conversion & 14.10\% & 15.22\% \\
										& Merge & SBOM aggregation & 11.54\% & 7.61\% \\
										& Tool Support & Integration support & 37.18\% & 52.17\% \\
										&	   & Total & 46.15\% & 63.04\% \\

			\bottomrule
		\end{tabular}
		\label{tab:tool_distribution}
	\end{center}
\end{table}
 
\begin{figure}[t]
    \centerline{\includegraphics[width=\linewidth]{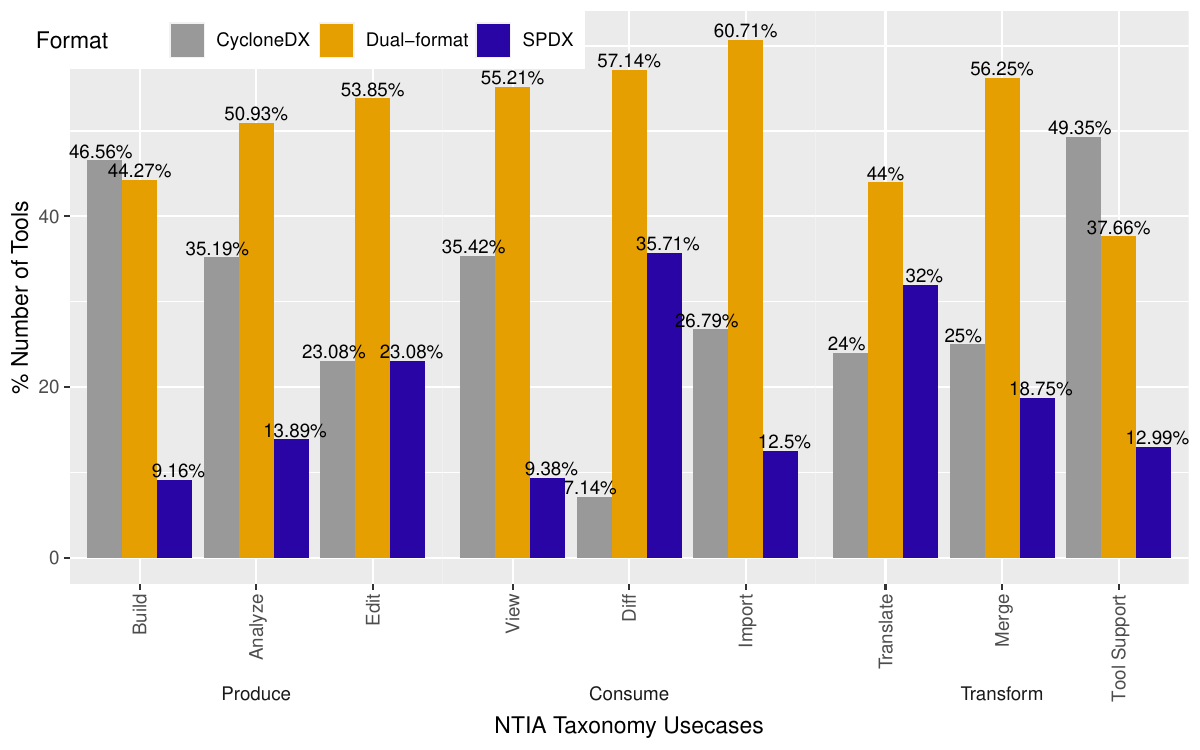}}
    \caption{Distribution of the 170 manually analyzed tools across the NTIA use cases, by supported format. Dual-format tools support both formats. For each use case, the percentages sum to 100\%.}
    \label{fig:ntia_spdx_cdx_dual}
\end{figure}

\textbf{\cdx-only tools are negatively associated with the \emph{Diff} use case, dual-format tools with the \emph{Tool Support} use case, and \spdx-only tools with the \emph{Build} use case. Conversely, \cdx-only tools are positively associated with the \emph{Build} and \emph{Tool Support} use cases, dual-format tools with the \emph{Import} use case, and \spdx-only tools with the \emph{Diff} and \emph{Translate} use cases.}
\tabref{tab:tool_distribution} presents the most prevalent use cases, based on the NTIA tool taxonomy, among single- and dual-format tools, while Figure~\ref{fig:ntia_spdx_cdx_dual} further separates the single-format tools into \spdx-only and \cdx-only groups.
Although the figure may give the impression that \spdx tools are less represented across all use cases, this interpretation can be misleading. In reality, dual-format tools are the most common in nearly all use cases, while \cdx has a larger number of tools with exclusive support than \spdx has. The smaller count of tools explicitly supporting only \spdx (19), compared to 73 for \cdx and 78 for dual-format, suggests that \cdx may lean more toward exclusivity, while many tools supporting \spdx also accommodate other formats.
A chi-square test, conducted at a significance level of $<0.05$, revealed a significant difference between the formats in terms of the use cases their tools support, with a small Cramér's V effect size of 0.181, interpreted with Cohen's benchmarks for $w = V\sqrt{\min(r-1,c-1)}$~\citep{effectSize}.
This indicates that the use cases supported by each format's tools differ, although the effect is small.

\begin{figure}[t]
    \centerline{\includegraphics[width=\linewidth]{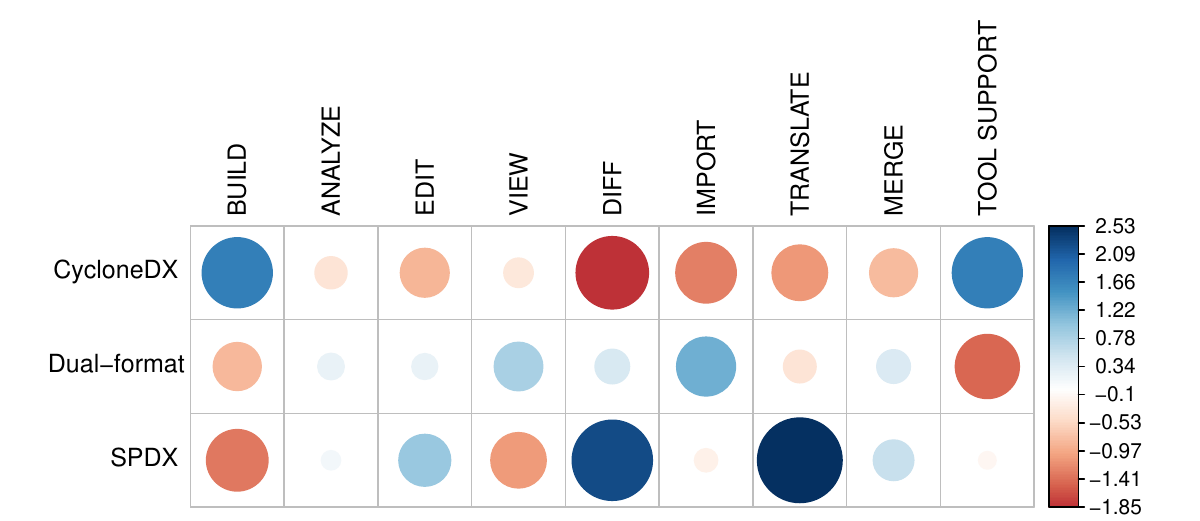}}
    \caption{Visualization of the Pearson residuals to compare the use cases of \spdx, \cdx, and dual-format SBOM tools. A positive residual value indicates a positive association, while a negative residual value indicates a negative association between an SBOM format and a use case. The size of the circle indicates strength of the association of an SBOM format with a use case.}
    \label{fig:cdx_vs_spdx_residuals}
\end{figure}

Since tools that support SBOM formats vary significantly in their coverage of use cases, we analyze the effect size using Cramér's V to identify which use cases are strongly or weakly associated with specific SBOM formats. To investigate the associations, we calculate the Pearson residuals of each observation from the chi-square test.
In this case, the Pearson residual highlights the strength of the association between a use case and an SBOM format. Only the \spdx-only residuals for \emph{Diff} ($2.275$) and \emph{Translate} ($2.533$) exceed $1.96$ in absolute value, so we read the other associations as descriptive tendencies.
We present the residual values in \tabref{tab:cdx_vs_spdx_residuals}, and a visual representation of the same data, for easier interpretation, in Figure~\ref{fig:cdx_vs_spdx_residuals}.

The figure shows that \cdx-only tools are negatively associated with the \emph{Diff} use case. Dual-format tools are negatively associated with the \emph{Tool Support} use case. Furthermore, \spdx-only tools are negatively associated with the \emph{Build} use case.
On the other hand, \cdx-only tools are positively associated with the \emph{Build} and \emph{Tool Support} use cases, dual-format tools are positively associated with the \emph{Import} use case, and \spdx-only tools are positively associated with the \emph{Diff} and \emph{Translate} use cases.

We would like to highlight that Microsoft has stated that it uses \spdx for all SBOMs it generates~\citep{diglio2021spdx}. Although our dataset indicates limited \emph{Build} support among \spdx-only tools, GitHub can export the dependency graph of a repository as an \spdx SBOM~\citep{githubsbomexport}. Thus, while \spdx may appear to have fewer associated build tools, a larger proportion of software may be built with \spdx as the SBOM format than these tool counts suggest.

\begin{table*}
	\centering
	\caption{The Pearson residuals and contribution proportion (\%) comparison between the use cases of \spdx, \cdx, and dual-format (DF) SBOM tools.}
	\begin{tabular}{llrrrrrr}
		\toprule
		Category & Use case & \spdx & \cdx & DF & \spdx~(\%) & \cdx~(\%) & DF~(\%)\\
		\cmidrule(lrr){1-8}
                \multirow{3}{*}{Produce} & Build & -1.334 & 1.733 & -0.812 & 5.054 & 8.530 & 1.872 \\
        & Analyze & 0.129 & -0.362 & 0.248 & 0.047 & 0.372 & 0.174 \\
       &  Edit & 0.949 & -0.84 & 0.236 & 2.555 & 2.005 & 0.158 \\
        \cmidrule(lrr){2-8}
                \multirow{3}{*}{Consume} & View & -1.085 & -0.304 & 0.831 & 3.341 & 0.263 & 1.962 \\
  &       Diff & 2.275 & -1.848 & 0.421 & 14.688 & 9.695 & 0.502 \\
     &    Import & -0.19 & -1.29 & 1.222 & 0.103 & 4.722 & 4.239 \\
        \cmidrule(lrr){2-8}
                \multirow{3}{*}{Transform} & Translate & 2.533 & -1.09 & -0.374 & 18.213 & 3.371 & 0.398 \\
        & Merge & 0.58 & -0.806 & 0.399 & 0.956 & 1.846 & 0.451 \\
      & Tool Support & -0.107 & 1.729 & -1.449 & 0.032 & 8.488 & 5.963 \\
		\bottomrule 
	\end{tabular}
	\label{tab:cdx_vs_spdx_residuals}
\end{table*}
 
In the previous chi-square test, we used Pearson's residual values to determine which use cases contributed the most to the overall statistic. Specifically, these values highlighted the observed differences in the frequency of use cases between tools supporting \spdx, \cdx, or both formats. As shown in \tabref{tab:cdx_vs_spdx_residuals}, the top three contributing use cases are \emph{Diff} (24.89\%), \emph{Translate} (21.98\%), and \emph{Build} (15.46\%), which together account for 62.32\% of the total chi-square value. This indicates that these use cases drive most of the variation across formats. Specifically, \spdx-only tools are positively associated with the \emph{Translate} use case and negatively associated with \emph{Build}. In contrast, \cdx-only tools are positively associated with \emph{Build} and negatively associated with \emph{Diff}. Dual-format tools are positively associated with \emph{Import} and negatively associated with the \emph{Tool Support} use case.

\summarybox{RQ1: Supported use cases of \spdx versus \cdx.}{lightgray}{white}{We find that the tools of the two SBOM formats exhibit distinct strengths and focus areas. These insights help SBOM tool adopters choose tools that align with their specific use cases, and guide SBOM tool developers in tailoring their tools to address gaps and reinforce existing strengths.}

\subsubsection{OSS versus Proprietary SBOM Tools.}

\begin{figure}[t]
    \centerline{\includegraphics[width=\linewidth]{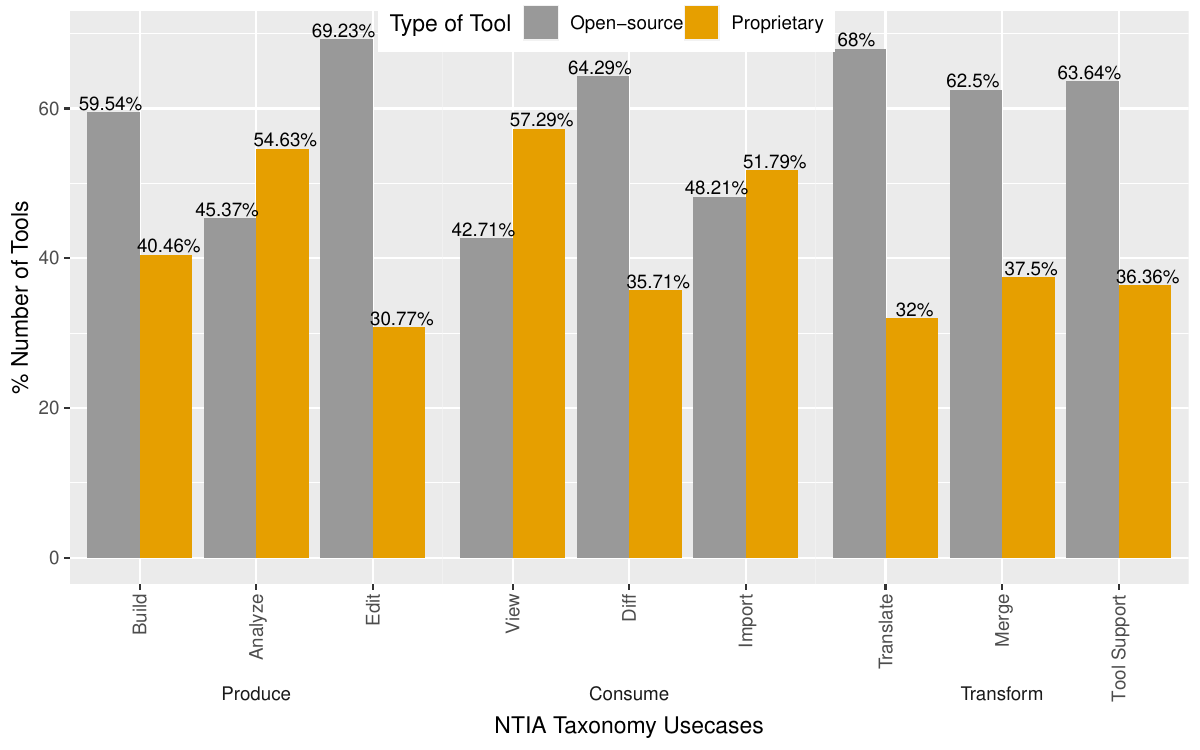}}
    \caption{Comparison of how $170$ SBOM tools (both OSS and proprietary) distributed across the NTIA tool taxonomy.}
    \label{fig:tool_classification}
\end{figure} \textbf{Proprietary SBOM tools across \spdx and \cdx support more use cases than the OSS SBOM tools.} We manually identify the NTIA tool taxonomy use cases supported by 108 OSS and 62 proprietary SBOM tools in the tool center, comparing their coverage to assess whether OSS tools are as comprehensive as proprietary ones. A chi-square test of independence reveals a statistically significant difference between the two, with a $p$-value of $<0.05$ and a small Cramér's V effect size of $0.182$~\citep{effectSize}.

\textbf{Proprietary tools are positively associated with the \emph{Analyze} and \emph{View} use cases, while OSS tools are negatively associated with them. Conversely, OSS tools are positively associated with the \emph{Tool Support} use case, while proprietary tools are negatively associated with it.} Given the significant difference in use case coverage, we further investigate the Cramér's V effect size to explore the strength of association between tool types and specific use cases. As previously explained, we calculate Pearson residuals to identify which factors most contribute to the chi-square test results, reflecting the strength of the association between use cases and tool types. No residual exceeds $1.96$ in absolute value (largest: $1.618$ for \emph{View}), so we read these associations as descriptive tendencies.

\begin{figure}[t]
    \centerline{\includegraphics[width=\linewidth]{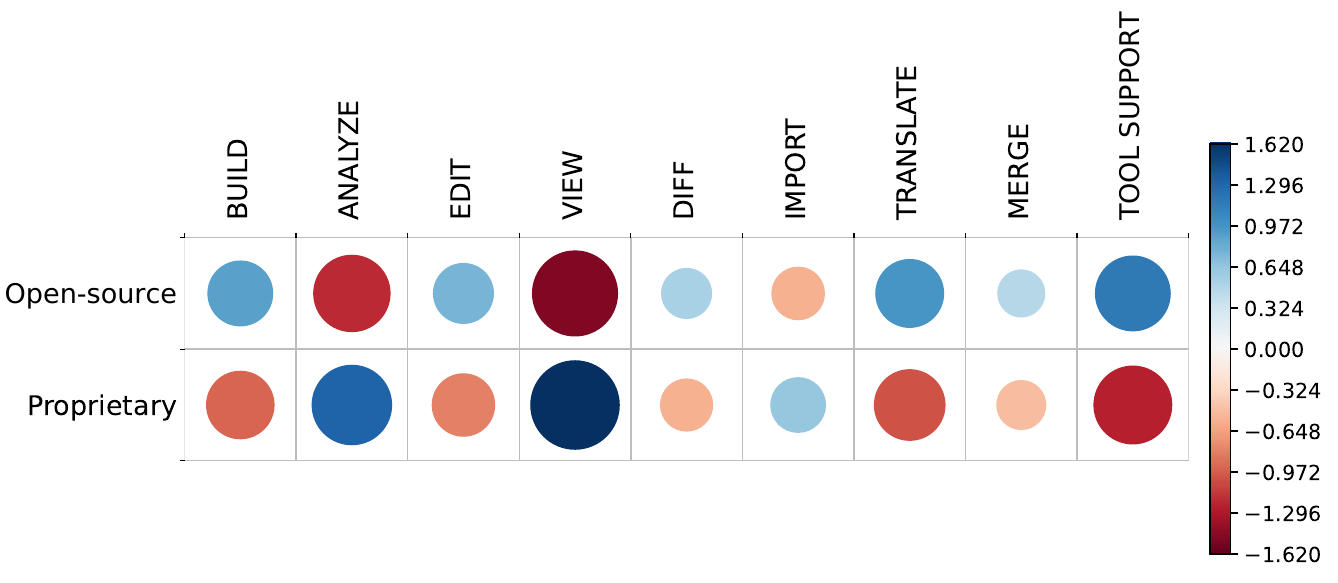}}
    \caption{Visualization of the Pearson residuals to compare the use cases of OSS versus Proprietary SBOM tools.}
    \label{fig:open_vs_prop_residuals}
\end{figure}

The residual values are presented in \tabref{tab:open_vs_prop_residuals} and visually represented in Figure~\ref{fig:open_vs_prop_residuals}. In the figure, negative residuals (in red) indicate a negative association, while positive residuals (in blue) indicate a positive association. The results show that OSS tools are negatively associated with the \emph{Analyze} and \emph{View} use cases, while they are positively associated with the \emph{Build}, \emph{Tool Support}, and \emph{Translate} use cases, with which proprietary tools are negatively associated. Proprietary tools are positively associated with \emph{Analyze} and \emph{View}.

The size of the circles in the figure reflects the contribution of each tool type to a given use case. To quantify these contributions, we calculate the residual values' proportions for each use case, presented in \tabref{tab:open_vs_prop_residuals}. Since these proportions are shares of the same total chi-square statistic, the last column of the table sums them across the two tool types. The top contributors are \emph{View} ($27.49\%$), \emph{Analyze} ($17.98\%$), and \emph{Tool Support} ($16.57\%$), which together account for $62.04\%$ of the total chi-square difference between OSS and proprietary tools.

\begin{table*}
	\centering
	\caption{The Pearson residuals and contribution proportion (\%) comparison between the use cases of all tool-center OSS versus proprietary SBOM tools. The contribution proportions of all 18 cells (9 use cases $\times$ 2 tool types) sum to $100\%$. The last column adds the proprietary and OSS proportions of each use case.}

	\begin{tabular}{llrrrrr}
		\toprule
        & & \multicolumn{2}{c}{Pearson Residuals} & \multicolumn{3}{c}{Contribution Proportion~(\%)} \\
		Category & Use case & Proprietary & OSS & Proprietary & OSS & Both\\

		\cmidrule(lrr){1-7}
	                \multirow{3}{*}{Produce} & Build & -0.948 & 0.877  & 5.091 & 4.351 & 9.442 \\
	        & Analyze & 1.309 & -1.210 & 9.695 & 8.286 & 17.981 \\
	       &  Edit & -0.813 & 0.752 & 3.746 & 3.201 & 6.947 \\
        \cmidrule(lrr){2-7}
	                \multirow{3}{*}{Consume} & View & 1.618 & -1.496 & 14.822 & 12.668 & 27.490 \\
	  &       Diff & -0.571 & 0.528 & 1.849 & 1.580 & 3.429 \\
	     &    Import & 0.629 & -0.581 & 2.238 & 1.913 & 4.151 \\
        \cmidrule(lrr){2-7}
	                \multirow{3}{*}{Transform} & Translate & -1.037 & 0.959 & 6.092 & 5.206 & 11.298 \\
	        & Merge & -0.506 & 0.468 & 1.448 & 1.238 & 2.686 \\
	      & Tool Support & -1.256 & 1.161 & 8.936 & 7.637 & 16.573 \\
		\bottomrule 
	\end{tabular}

	\label{tab:open_vs_prop_residuals}
\end{table*}
 
In general, despite being fewer in number compared to OSS tools, proprietary SBOM tools support an average of $3.98$ use cases per tool versus $2.68$ for OSS tools, or $1.49$ times as many. Our analysis reveals that \emph{View} has the largest positive residual for proprietary tools ($1.618$), while \emph{Tool Support} has the largest positive residual for OSS tools ($1.161$). Conversely, proprietary tools are negatively associated with \emph{Tool Support}, and OSS tools are negatively associated with \emph{View}.

\summarybox{RQ1: OSS versus Proprietary contribution towards use cases.}{lightgray}{white}{
    Our results show that supported use cases are unevenly distributed between OSS and proprietary tools. The OSS community could prioritize developing visualization tools to support decision-making, while proprietary efforts might focus on expanding libraries and APIs to improve integration capabilities.}
 \subsection{State of the OSS SBOM tools (RQ2)}

In RQ2, we inspect the community health of the OSS \spdx and \cdx GitHub repositories in our dataset. This dataset comprises repositories identified from the official tool centers and contributing forks identifiable during the collection window. We analyze important CHAOSS community-health metrics\footnote{https://chaoss.community/kbtopic/all-metrics}, such as the number of contributors, commits, forks, stars, watchers, PRs, and releases, and the rate of returning contributors. This analysis reveals the level of community engagement, activity, and sustainability of the sampled OSS tools, which could be important indicators to use before selecting a tool for one's project.

\begin{figure}[t]
    \centerline{\includegraphics[width=\linewidth]{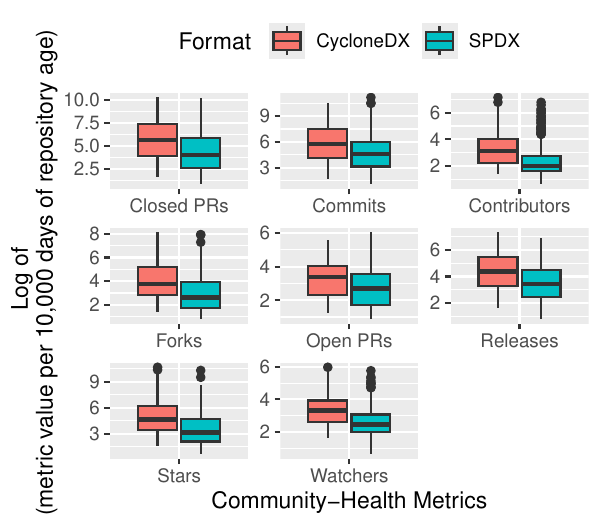}}  \caption{Comparison of the distribution of normalized metrics (by repository age), i.e., \#PRs, \#Contributors, \#Stars, \#Watchers, \#Releases, \#Forks representing the community-health metrics of 456 \spdx and 145 \cdx tool repositories.}
    \label{fig:tool_metrics}
\end{figure}
 \begin{table*}
	\centering
	\caption{Mann-Whitney U test~\citep{mann1947test} applied to evaluate the difference between normalized community-health metrics of the 145 \cdx tool-repositories and 456 \spdx tool-repositories. $p$-values are Bonferroni-corrected over the eight metrics, and Cliff's delta is positive when \cdx values exceed \spdx values. $p < 0.05$ indicates a significant difference between the two distributions.}

	\begin{tabular}{lrrl}
		\toprule
		Metric & p-value  &Cliff's delta & Effect Size~\citep{romano2006appropriate}\\
		\cmidrule(lrr){1-4}
		ClosedPullRequest & $1.70\times10^{-10}$ & $0.351$ & Medium \\
        Commits & $4.44\times10^{-06}$ & $0.280$ & Small \\
		Contributors & $5.12\times10^{-14}$ & $0.435$ & Medium \\
		Forks & $8.33\times10^{-10}$ & $0.349$ & Medium \\
		OpenPullRequest & $1.15\times10^{-09}$ & $0.300$ & Small \\
		Releases & $6.96\times10^{-09}$ & $0.279$ & Small \\
		Stars & $1.13\times10^{-11}$ & $0.382$ & Medium \\
		Watchers & $2.48\times10^{-14}$ & $0.441$ & Medium \\
		\bottomrule
	\end{tabular}

	\label{tab:metric_difference}
\end{table*}
 
\subsubsection{Insights on the community-health metrics of the sampled tool repositories.}

\textbf{All OSS community-health metrics are significantly higher for \cdx than for \spdx, with a small to medium Cliff's delta effect size.}
To assess the overall health and activity of the identified tool repositories, we collect key CHAOSS community-health metrics, including the number of stars, forks, releases, contributors, commits, watchers, and the counts of open/closed PRs. 
Since every repository has a different maturity level, we normalize the collected values of the metrics by the age of the repository (per 10,000 days).
These metrics are commonly used to evaluate activeness, maturity, maintenance efforts, and user engagement in OSS communities~\citep{zerouali2019diversity, foundjem2021onboarding, gao2024characterizing}.

\figref{fig:tool_metrics} illustrates the results from the OSS GitHub tool repositories of \spdx and \cdx tools. To compare the difference across the metrics between the tools of the two formats, we apply the Mann-Whitney U test~\citep{mann1947test} to the normalized community-health metrics, followed by a Bonferroni correction to mitigate the risk of false positives. The results of this test are presented in \tabref{tab:metric_difference}, with $p<0.05$ indicating a significant difference and Cliff's delta~\citep{cliff1993dominance} values indicating the effect size of this difference, which we interpret with the thresholds of Romano et al.~\citep{romano2006appropriate} (negligible below $0.147$, small below $0.33$, medium below $0.474$, and large otherwise).

These results show significant differences in all community-health metrics between the two formats, with Cliff's delta indicating small to medium differences (0.28 for Releases to 0.44 for Watchers). Given the unequal number of tool repositories in each format, we performed $1{,}000$ bootstrap samples~\citep{efron1979bootstrap} with a $95\%$ confidence interval to determine the true median for each metric. We use this median value in the textual discussion of RQ2. Next, we look deeper into the different CHAOSS community-health metrics\footnote{https://chaoss.community/kbtopic/all-metrics} that were studied.

\cite{borges2018s} conducted an empirical study on GitHub projects, identifying three main reasons why developers star a project: to express appreciation, to bookmark the project for later use, or since they are actively using it. They also find that $75\%$ of developers consider the number of stars before using or contributing to a project. Based on this, they suggest that project owners monitor their star count relative to competitors to gauge their project's attractiveness to the OSS community. 

When comparing \cdx and \spdx tools, \cdx tools have a statistically higher median normalized star count ($46.9$) compared to \spdx tools ($4.45$). While this suggests that \cdx tools may have garnered more attention from the OSS community, the difference could be influenced by factors such as the number of contributors or the broader use of the tools in specific domains. Therefore, the higher star count does not necessarily indicate a greater likelihood of contributions but may reflect other factors such as visibility or project maturity. Moreover, developers or users watch a repository to receive notifications about its activity, such as new issue reports, PRs, or discussions. Similar to stars, the number of watchers reflects the level of engagement within a project.

Despite \cdx having fewer OSS tool repositories than \spdx, its median number of watchers ($27.7$) exceeds that of \spdx ($11.5$). This suggests that \cdx tools attract more attention from observers in the OSS community~\citep{borges2018s}. The ``fork'' metric, on the other hand, measures how many copies of a project exist on a code development platform~\citep{chaoss}. In this case, \cdx tools again outperform \spdx, with a median number of normalized forks ($20.8$) significantly higher than that of \spdx ($2.93$). This indicates greater interest in \cdx tools, not only in terms of observation but also in terms of users creating personal copies for further development.

\textbf{\cdx tools have more PRs, releases, contributors, and contributions than \spdx.} PRs are a common mechanism for proposing changes to source code repositories, enabling stakeholders to review and discuss modifications before integration~\citep{chopra2021alex}. \cdx surpasses \spdx in both the median and mean number of normalized open and closed PRs, suggesting a higher level of ongoing development and code review activity. Specifically, \cdx has a median of 5.56 open PRs (mean: 22.3), whereas \spdx has a median of 0 (mean: 8.27). Similarly, the number of normalized closed PRs in \cdx (median: 54, mean: 1,124) significantly exceeds \spdx (median: 0, mean: 338).

A similar trend is observed in commit activity: \cdx tools have a higher median (324) and mean (1,973) number of normalized commits compared to \spdx tools (median: 100, mean: 1,055), indicating more frequent updates and iterative improvements. The number of normalized project releases serves as an indicator of development progress, update frequency, and versioning practices. Here too, \cdx exhibits higher values (median: 0, mean: 98.2) than \spdx (median: 0, mean: 21.3), implying more active release cycles.

Finally, contributor count reflects the collaborative nature and community engagement of a project. In our analysis, \cdx tools show higher normalized contributor counts (median: 21.4, mean: 72.2) compared to \spdx tools (median: 7.29, mean: 28.3), suggesting broader community involvement. While these findings represent a snapshot in time, they underscore that choosing an SBOM format should involve not only evaluating the specification itself but also considering the level of activity and community dynamics. Future work could investigate qualitative factors, such as governance models, funding sources, or adoption trajectories, that may explain the observed differences in development engagement between \cdx and \spdx.

\summarybox{RQ2: Community health of tools}{lightgray}{white}{The number of normalized stars, watchers, and forks suggests that, despite having fewer tool repositories than \spdx, the OSS community is more engaged with \cdx tools. The higher star number of these tools is consistent with OSS developers preferring to contribute to \cdx tools over \spdx tools~\citep{borges2018s}, although stars alone do not establish this preference. Similarly, across PRs, commits, contributors, and releases, \cdx tools exhibit greater development activity than \spdx tools, despite their smaller set of tool repositories.
One hypothesis is that this engagement relates to the use cases that \cdx tools support (which differ from \spdx as shown in RQ1), but we have not tested it.
}

\begin{table*}
\centering
\caption{Result of Mann-Whitney U test to compare the difference between returning contributor groups for \spdx and \cdx tool repositories. Cliff's delta is positive when \spdx values exceed \cdx values. Each group includes its lower bound and excludes its upper bound, and significance is assessed after Bonferroni correction ($\alpha=0.01$).}
\label{tab:returning_commit_count_distribution}
\begin{tabular}{lrrlr}
\toprule
Groups & p-value & Cliff's delta & Effect size & Significance \\
\midrule
2-10 & 0.015 & 0.066 & negligible & False \\
10-100 & 0.546 & -0.025 & negligible & False \\
100-1,000 & 0.200 & -0.093 & negligible & False \\
1,000+ & 0.894 & -0.045 & negligible & False \\
all & $<0.001$ & 0.096 & negligible & True \\
\bottomrule
\end{tabular}
\end{table*}
 \begin{figure}[t]
    \centerline{\includegraphics[width=1.2\linewidth]{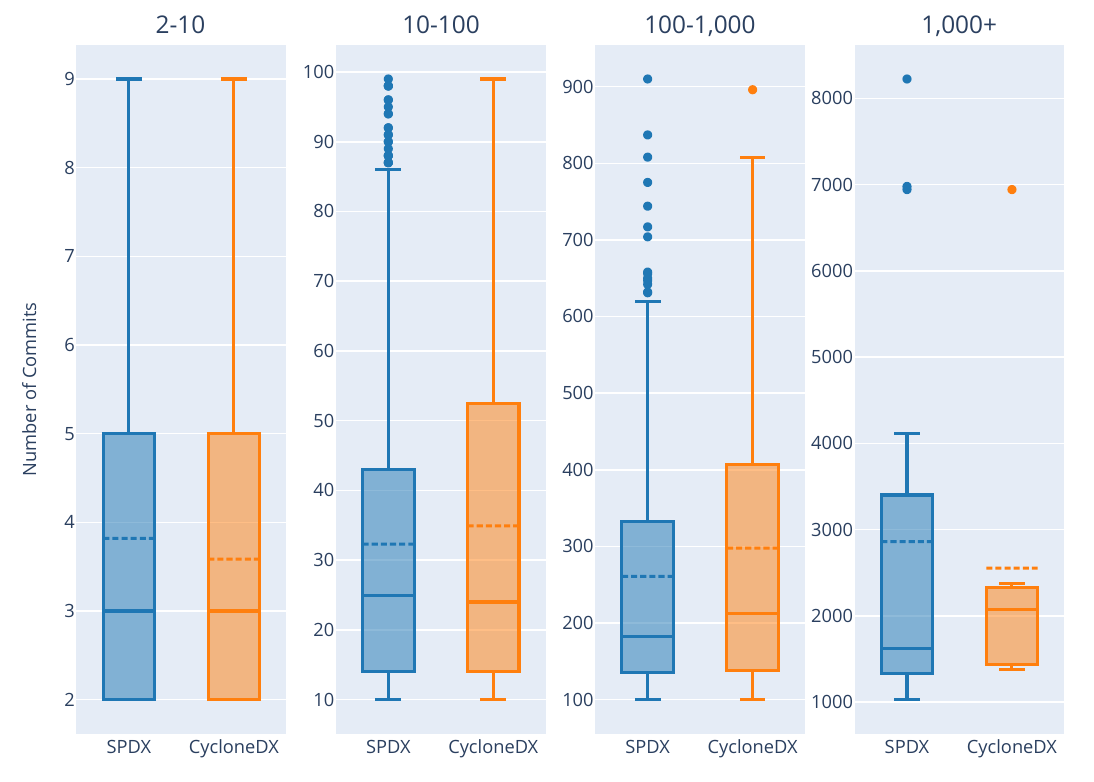}}
    \caption{Distribution of returning contributor groups (by number of commits) across $456$ \spdx and $145$ \cdx tool repositories.}
    \label{fig:returning_commit_count_distribution}
\end{figure}
 
\textbf{\spdx tools have a higher rate of returning contributors than \cdx tools, but the number of commits per returning contributor differs negligibly.}
A returning contributor of a project is a GitHub contributor who contributes to a project more than once. Counting each contributor once per repository and format, $1{,}838$ of $2{,}940$ \spdx contributors ($62.5\%$) and $1{,}092$ of $2{,}103$ \cdx contributors ($51.9\%$) are returning contributors, a significant difference with a small effect size ($\chi^2=56.49$, $p<0.001$, $\phi=0.106$). We then group the returning contributors by their number of commits and compare the groups between the two formats (\spdx and \cdx).
\tabref{tab:returning_commit_count_distribution} presents the Mann-Whitney U test statistics for the difference of \spdx and \cdx tools across the returning contributor groups (by number of commits), including p-values, Cliff's delta, effect sizes, and significance after Bonferroni correction ($\alpha=0.01$ over five tests). 

We categorized returning contributors into four groups based on their number of commits to a repository ($2$--$10$, $10$--$100$, $100$--$1{,}000$, and $1{,}000+$ commits), where each group includes its lower bound and excludes its upper bound. For the low commit count group ($2$--$10$ commits), the p-value is $0.015$, which is not significant after correction, and the effect size (Cliff's delta) of $0.066$ is negligible. For the medium commit count group ($10$--$100$ commits), the p-value is $0.546$, showing no significant difference. Similarly, for the high commit count group ($100$--$1{,}000$ commits) and the extremely high commit count group ($1{,}000+$ commits), the p-values are $0.200$ and $0.894$, respectively, both indicating no significant difference. When considering all commit count groups combined, the p-value is $<0.001$, showing a significant difference, but the effect size remains negligible at $0.096$.

\figref{fig:returning_commit_count_distribution} shows the distribution of returning contributors across various commit count groups for \spdx and \cdx tool repositories. The median and mean number of commits for \spdx ($3$ and $3.82$) and \cdx ($3$ and $3.58$) are similar for low commit counts ($2$--$10$). Similarly, for medium commit counts ($10$--$100$), both \spdx (median: $25$, mean: $32.31$) and \cdx (median: $24$, mean: $34.91$) exhibit a comparable level of contribution among returning contributors. For high commit counts ($100$--$1{,}000$), \cdx (median: $212$, mean: $297.87$) shows a slightly higher level of contribution than \spdx (median: $183$, mean: $260.74$). Finally, for extremely high commit counts ($1{,}000+$), \spdx has a lower median but a higher mean contribution level (median: $1{,}624$, mean: $2{,}860.38$) than \cdx (median: $2{,}071$, mean: $2{,}554$).

\summarybox{RQ2: Returning Contributors}{lightgray}{white}{\spdx tools have a higher rate of returning contributors than \cdx tools ($62.5\%$ versus $51.9\%$, small effect). Among returning contributors, the number of commits per contributor differs negligibly between the formats in every commit count group.}
 \subsection{Issue reports in the OSS SBOM tools (RQ3)}
\label{sec:rq3-result}

Since previous studies have demonstrated the potential of mining GitHub issue reports to uncover developers' concerns and challenges~\citep{kikas2015issue,tan2020first}, this RQ examines the prevalence of GitHub issue reports in the OSS tool repositories that support the \cdx and \spdx formats. Issue reports in GitHub repositories serve as a means to report, discuss, and track various aspects of development. Additionally, insights into the nature of these issue reports can aid SBOM practitioners in navigating common pain points and making informed decisions about adoption and implementation within their supply chains. Moreover, a deeper understanding of issue patterns would allow developers to address recurring challenges, ultimately contributing to more robust and sustainable SBOM tools. Understanding the resolution time for different issue categories can shed light on the complexity of specific SBOM-related challenges, helping to identify particularly difficult use cases and tools that require further attention.

\subsubsection{Comparing closed and open issue reports.}

\begin{figure}[t]
    \centerline{\includegraphics[width=\linewidth]{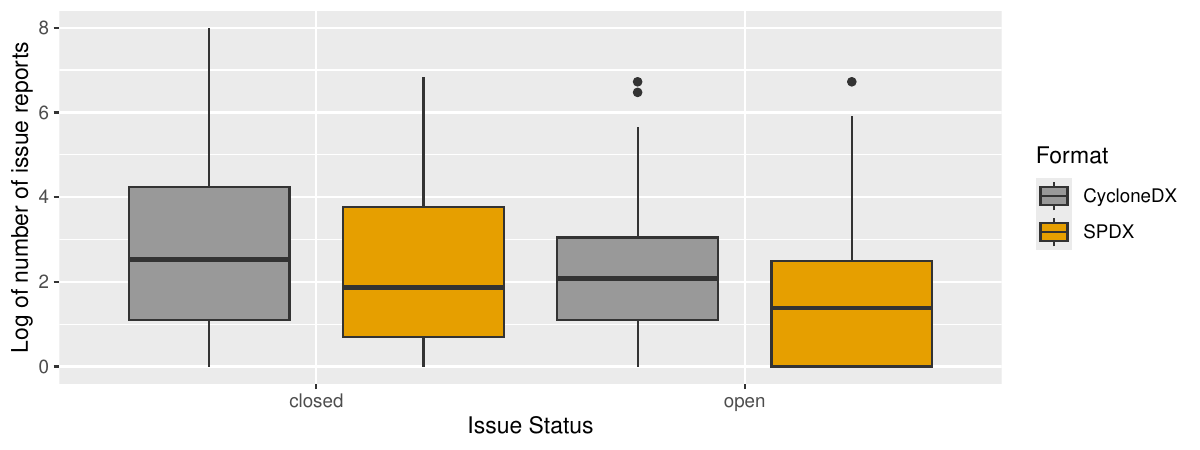}}
    \caption{Representation of 14,811 issue reports in 90 \cdx tool repositories and 12,042 issue reports in 171 \spdx tool repositories (created since 2018 and counted once per format).}
    \label{fig:issue_stats}
\end{figure}

\textbf{\cdx tools have a higher median number of issue reports per repository than \spdx tools.} The number of issue reports, whether open or closed, can indicate the level of activity, responsiveness to bug reports or feature requests, and overall project health~\citep{kikas2015issue}. Additionally, the more a project is used, the more issue reports are likely to be filed~\citep{6079845}. We compare the \cdx and \spdx tool repositories by their number of issue reports per repository using the Mann-Whitney U test~\citep{mann1947test} and find a statistically significant difference between the tools of both formats, with $p=0.0017$ and a small Cliff's delta effect size of $0.236$ (median of 15 issue reports per \cdx repository versus 7 per \spdx repository).
For a fair comparison, in this RQ, we consider only those issue reports that were created since 2018 because \cdx technology was released in May 2017. We compare raw per-repository counts and do not normalize them by repository age.

\textbf{\cdx tools have a higher mean number of issue reports than \spdx tools, but the distributions differ significantly only for open issue reports.}
Across the open and closed repository-status groups, the mean number of issue reports is higher for \cdx than for \spdx, yet a $1{,}000$-sample bootstrap~\citep{efron1979bootstrap} with a 95\% confidence interval shows that the formats' intervals overlap (\cdx: 38.9--134.3, \spdx: 27.3--59.8), so the bootstrap does not establish a difference in means.
We then group the issue reports by status. \figref{fig:issue_stats} shows 3,504 open issue reports in 171 \spdx tool repositories compared to 3,729 in 90 \cdx tool repositories, and 8,538 closed issue reports in \spdx tool repositories compared to 11,082 in \cdx tool repositories.
The mean number of open issue reports per repository is 45.48 in \cdx (standard deviation 126.37) versus 27.16 in \spdx (88.86), and the mean number of closed issue reports is 142.08 (415.74) versus 61.87 (160.06).
We apply the Mann-Whitney U test to the open and closed issue reports per repository (a repository counts in a status group if it has at least one issue report of that status).
For open issue reports, we find a statistical difference with $p=0.019$ and a small Cliff's delta of $0.191$, whereas for closed issue reports the difference is not significant at the 5\% level ($p=0.055$, Cliff's delta $0.157$, small).
The higher standard deviation of \cdx in both statuses indicates that issue activity is more uneven across \cdx tools, with a few very active repositories driving the mean.
These counts are not normalized by repository age, so the age of the repositories may partly explain them. One hypothesis is that the younger and faster-growing set of \cdx tools attracts more new issue reports, but our data do not test this.

\summarybox{RQ3: Issue Activity}{lightgray}{white}{\cdx tool repositories receive more issue reports than \spdx tool repositories (median 15 versus 7 per repository, small effect), with the difference significant for open but not for closed issue reports. These differences are associated with the format that a tool supports, which adopters may therefore consider when selecting SBOM tools, although our data do not show that the format or its tooling community causes them.}

\begin{figure}[t]
    \centerline{\includegraphics[width=\linewidth]{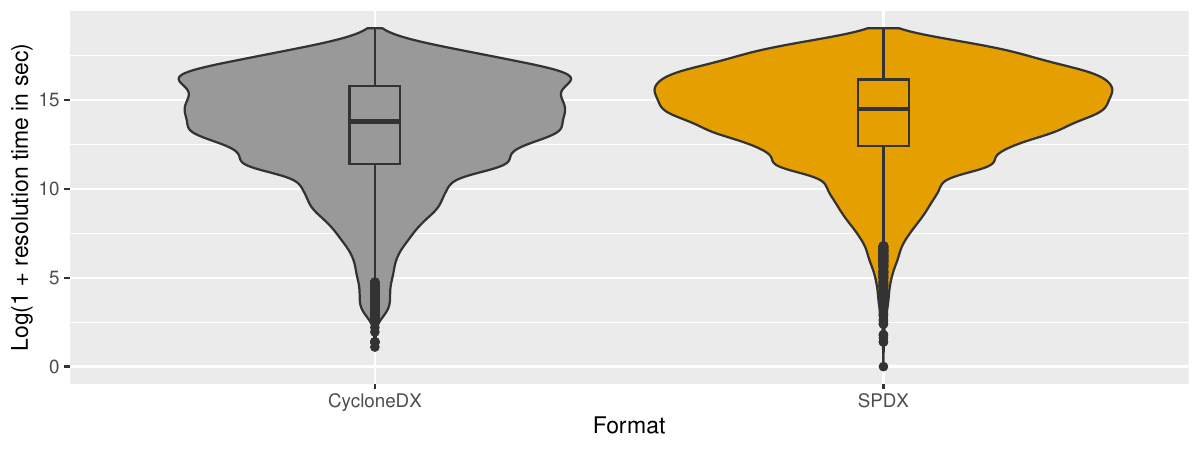}}
    \caption{Resolution time of 11,082 closed issue reports in 78 \cdx tool repositories and 8,538 closed issue reports in 138 \spdx tool repositories (created since 2018 and counted once per format).}
    \label{fig:resolve_time}
\end{figure}

\subsubsection{Comparing resolution time of the issue reports.}

\textbf{\cdx tools have a faster and more stable resolution time for their GitHub issue reports than \spdx, although the effect is negligible in size.}
We investigate if the tools of one format have a shorter resolution time for the issue reports when compared to the tools of another format.
To compare the resolution time across \cdx and \spdx, we apply the Mann-Whitney U test and find a significant difference between the two ($p = 2.1\times10^{-53}$) with a negligible Cliff's delta of $0.128$. \figref{fig:resolve_time} presents the resolution times, showing that despite having a higher number of issue reports than \spdx, \cdx tools still achieve a lower resolution time. On average, \spdx issue reports take 121.23 days to resolve, whereas \cdx issue reports are resolved in 85.36 days, a $29.6\%$ reduction (median: 22.1 versus 11.1 days). Additionally, the standard deviation of issue resolution time is 237.1 days for \spdx tools versus 178.7 days for \cdx tools (33\% higher), indicating notable variability in how the tools of different SBOM formats handle issue reports~\citep{kikas2015issue}.

\begin{figure}[t]
    \centering
    \begin{subfigure}{0.48\linewidth}\centering
        \includegraphics[width=\linewidth]{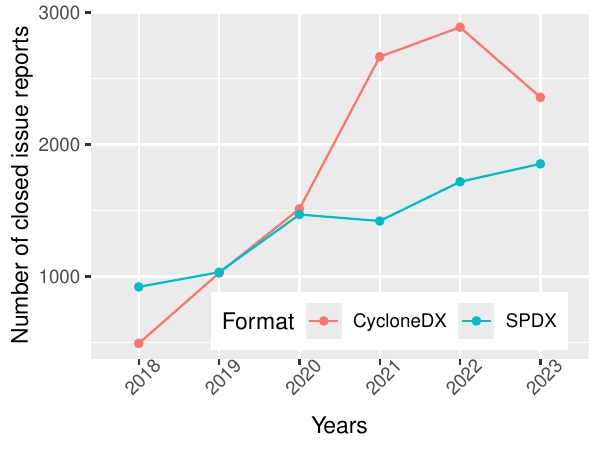}
        \caption{Prevalence of issue reports.}
    \end{subfigure}
    \hfill
    \begin{subfigure}{0.48\linewidth}\centering
        \includegraphics[width=\linewidth]{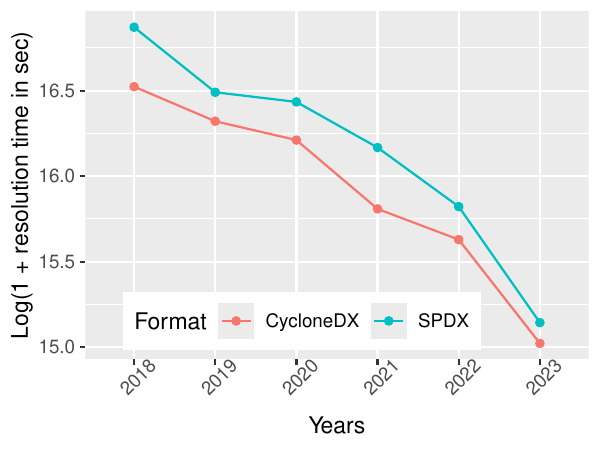}\caption{Resolution time of issue reports.}
    \end{subfigure}
    \caption{Trend of prevalence and resolution time of issue reports from 2018 to 2023.}
    \label{fig:trend_of_issues}
\end{figure}

To investigate the disparity in issue reporting between \spdx and \cdx tools from 2018 to 2023, we analyzed trends in closed issue reports. \figref{fig:trend_of_issues} shows (\spdx versus \cdx) that \spdx had more closed issue reports in 2018 (921 versus 493), the two formats were close in 2019 (1,032 versus 1,026) and 2020 (1,469 versus 1,512), and \cdx exceeded \spdx from 2021 onwards (1,420 versus 2,664 in 2021), while \cdx showed shorter mean resolution times in every year. The difference is significant after Bonferroni correction over the six years from 2020 onwards (adjusted $p=0.037$ in 2020 and $p<0.001$ from 2021), but not in 2018 ($p=0.85$) or 2019 ($p=0.14$), and the effect sizes are negligible to small ($|\delta|\le0.16$). Since this analysis only covers closed issue reports, later years had less time to close issue reports, which partly explains why the resolution times of both formats decline over time. The growth in \cdx closed issue reports coincides with the 2021 US executive order that required SBOMs for software sold to the federal government~\citep{eo14028}. Our data show the timing of this shift but cannot attribute it to the executive order.

\summarybox{RQ3: Issue Resolution}{lightgray}{white}{\cdx tools resolve issue reports faster on average (85.4 versus 121.2 days) and more consistently than \spdx tools, although the effect size is negligible, and \cdx tools have shorter mean resolution times in every year from 2018 to 2023, although significantly so after correction only from 2020 onwards.}

\subsubsection{Categorization of issue reports.}

Understanding the type of issue reports in each SBOM format would help tool developers improve their tools and help SBOM users make the right decision when selecting an SBOM tool for their project. We identify $14$ distinct categories by analyzing the tagged issue reports from the $91$ \spdx and $65$ \cdx tool repositories that have tagged issue reports (Section~\ref{sec:method:tag-grouping}), then apply the LLM-assisted approach (Section~\ref{sec:rq3-llm-classify}) to assign categories across the dataset. We have listed the indices and names of each category, accompanied by a short description, the set of relevant tags, and one representative example as follows.

\begin{longtable}{p{\linewidth}}
    \toprule
    ${C}_{01}$ \textbf{Bug Fixes and Defects} \\ 
    \midrule
    \textbf{Definition}: This category refers to technical problems within the codebase \\
    \textbf{Tags}: exceptions, stale, false-positive, defect, bug, kind/bug, technical issue, stale, good-first-issue, lifecycle/stale, lifecycle/frozen, QA \\
    \textbf{Example}: In kyverno issue \#9484\footnote{\url{https://github.com/kyverno/kyverno/issues/9484}}, there is a bug where Kyverno's \texttt{PolicyException} mechanism fails to recognize and exempt EphemeralContainers (used for debugging running pods) from policy enforcement, despite having proper exceptions configured for the namespace - affecting the debugging workflow in environments using Pod Security Standards policies. \\
    \midrule
    ${C}_{02}$ \textbf{Code Quality} \\ 
    \midrule
    \textbf{Definition}: This category refers to improving the overall quality of code. \\
    \textbf{Tags}: refactor, comment-styles, log enhancement, changelog-ignore, package-formats, package scan, performance, validation, generation \\
    \textbf{Example}: In tern issue \#948\footnote{\url{https://github.com/tern-tools/tern/issues/948}}, there is a significant refactoring effort of the project's container image handling architecture to replace ORAS with Skopeo for pulling container images. To improve performance and enhance validation, the developers create a new OCIImage class to handle different package formats (OCI versus Docker layouts), refactor the ImageLayer class with improved directory path generation methods, and validate compatibility across multiple container runtime interfaces to ensure better code quality and maintainability. \\
    \midrule
    ${C}_{03}$ \textbf{Community Engagement and Support} \\ 
    \midrule
    \textbf{Definition}: This category refers to engaging the community, seeking feedback, and providing support to contributors. \\
    \textbf{Tags}: request for comment, upstream, help wanted \\
    \textbf{Example}: In specification issue \#226\footnote{\url{https://github.com/CycloneDX/specification/issues/226}}, there is a community-driven request for comment emerging from the CycloneDX Attestations working group to enhance the specification by adding \texttt{bom-refs} (Bill of Materials references) to \texttt{OrganizationalEntity} and \texttt{OrganizationalContact} objects. The proposal seeks feedback from the broader community on introducing a new external reference type to support organizational referencing, representing collaborative upstream engagement where working group members are actively seeking support and input from contributors to refine the specification's organizational data model. \\
    \midrule
    ${C}_{04}$ \textbf{Configuration} \\ 
    \midrule
    \textbf{Definition}: This category refers to configuring, packaging, and managing dependencies for the project. \\
    \textbf{Tags}: configuration, docker, helm, installation and packaging, dependencies \\
    \textbf{Example}: In scancode-toolkit issue \#3210\footnote{\url{https://github.com/nexB/scancode-toolkit/issues/3210}}, there is a dependency management and packaging configuration challenge in ScanCode where native C-based libraries (\texttt{pyahocorasick}, \texttt{intbitset}, \texttt{lxml}) create portability issues across different environments. The proposed solution involves configuring fallback pure-Python alternatives for these dependencies to enable installation without native toolchain requirements, improving cross-platform packaging compatibility and supporting deployment in constrained environments such as browsers through better dependency configuration management. \\
    \midrule
    ${C}_{05}$ \textbf{Continuous Integration and Deployment (CI/CD)} \\ 
    \midrule
    \textbf{Definition}: This category refers to setting up and maintaining continuous integration and continuous deployment pipelines. \\
    \textbf{Tags}: CI, Type: CI, CI/CD \\
    \textbf{Example}: In schema issue \#174\footnote{\url{https://github.com/linterhub/schema/issues/174}}, there is a deployment pipeline issue where the GitHub Pages (\texttt{gh-pages}) automated deployment contains a low-quality logo that needs to be replaced with a high-resolution version, requiring updates to the CI/CD workflow that builds and deploys the static documentation site to ensure proper asset management and quality control in the continuous deployment process. \\
    \midrule
    
    ${C}_{06}$ \textbf{Documentation} \\ 
    \midrule
    \textbf{Definition}: This category refers to creating and maintaining documentation for the project. \\
    \textbf{Tags}: docs, Documentation, documentation \\
    \textbf{Example}: In cyclonedx-node-module issue \#234\footnote{\url{https://github.com/CycloneDX/cyclonedx-node-module/issues/234}}, there is a documentation issue where GitHub fails to properly detect and display the Apache 2.0 license for the CycloneDX Node.js module repository, preventing the automatic license summary from appearing at the top of the repository page and making it unavailable through GitHub's API for license detection tools. \\
    \midrule

    ${C}_{07}$ \textbf{Feature Development and Enhancement} \\ 
    \midrule
    \textbf{Definition}: This category refers to adding new features and enhancements to the project. \\
    \textbf{Tags}: kind/feature, proposed core enhancement, should have, nice to have, feature, must have, proposal, enhancement, enhancement, new feature, security \\
    \textbf{Example}: In sbom-operator issue \#172\footnote{\url{https://github.com/ckotzbauer/sbom-operator/issues/172}}, there is a feature request to enhance the sbom-operator by adding functionality to map Kubernetes pod labels as project tags in DependencyTrack, enabling better grouping and filtering capabilities based on application metadata like stage, department, and environment. \\
    \midrule
    ${C}_{08}$ \textbf{Integration and Interfacing} \\ 
    \midrule
    \textbf{Definition}: This category refers to integrating with various operating systems and ecosystems. \\
    \textbf{Tags}: os support, ecosystem:javascript, ecosystem:java, webhook \\
    \textbf{Example}: In scancode-toolkit issue \#2752\footnote{\url{https://github.com/nexB/scancode-toolkit/issues/2752}}, there is an integration issue with scancode-toolkit-mini's Python packaging where the tool fails to properly import the required \texttt{ctypes.util} module on newer Python versions, causing runtime failures when trying to execute the toolkit. The problem stems from incomplete dependency management in the mini distribution and missing explicit imports that newer Python ecosystems require, affecting the tool's ability to integrate seamlessly across different Python environments and operating system configurations where libmagic dependencies may be installed in non-standard locations. \\
    \midrule
    ${C}_{09}$ \textbf{Libraries} \\ 
    \midrule
    \textbf{Definition}: This category refers to usage of libraries and APIs. \\
    \textbf{Tags}: spdx-utils, API Call, core and api \\
    \textbf{Example}: In ort issue \#4740\footnote{\url{https://github.com/oss-review-toolkit/ort/issues/4740}}, there is an API usage issue where ORT's SPDX-utils library fails to properly integrate custom license texts through the core API, despite following the documented configuration approach with custom license directories and curation files. The problem involves the reporter tool's API calls not correctly accessing the configured custom license text files, indicating a gap in how the core and API components handle non-standard SPDX license text retrieval and integration into generated reports. \\
    \midrule
    ${C}_{10}$ \textbf{Licensing} \\ 
    \midrule
    \textbf{Definition}: This category refers to copyrights and licensing. \\
    \textbf{Tags}: improve-license-detection, copyright scan, Profile:Licensing, license-review, improve-license-detection, Submit New License, new license/exception: Not Accepted, new license/exception: Accepted, new license/exception request \\
    \textbf{Example}: In license-list-XML issue \#1365\footnote{\url{https://github.com/spdx/license-list-XML/issues/1365}}, there is a new license request for the ``Azratek Free and Open Source Software License'' (AzFOSS) submitted by Azratek, requiring license review and evaluation for inclusion in the SPDX License List. The submitted license comprises a standard license header with copyright attribution and warranty disclaimer provisions, requiring assessment against SPDX license inclusion criteria to determine whether it should be accepted or rejected for official recognition in license detection systems. \\
    \midrule
    ${C}_{11}$ \textbf{Code Components} \\ 
    \midrule
    \textbf{Definition}: This category refers to specific components of the codebase, such as serialization, parsers, and package scanning. \\
    \textbf{Tags}: serialization, data model, parser, analyzer, scanner, I/O, package-at-toplevel, package-files, package-formats, package scan, component:dep-sources \\
    \textbf{Example}: In scancode-toolkit issue \#2422\footnote{\url{https://github.com/nexB/scancode-toolkit/issues/2422}}, there is a packaging and integration issue with ScanCode's DWARF analyzer component that exists only in development branches but is not properly integrated into the main release distribution. The problem affects the binary analysis scanner's ability to parse DWARF debugging information from compiled code, requiring proper serialization of the plugin into PyPI packages and correction of I/O documentation to enable users to access this specialized parser for analyzing package files containing debugging symbols. \\
    \midrule
    ${C}_{12}$ \textbf{Release Management and Versioning} \\ 
    \midrule
    \textbf{Definition}: This category refers to managing releases, handling breaking changes, and versioning of the project. \\
    \textbf{Tags}: breaking change, release-high, pending release, breaking-change, breaking change \\
    \textbf{Example}: In syft issue \#1864\footnote{\url{https://github.com/anchore/syft/issues/1864}}, there is a breaking change planned for Syft's 1.0 release that involves removing deprecated configuration options including the \texttt{--name} flag and \texttt{Application.Name} configuration, representing a coordinated release management effort to clean up the CLI while maintaining backwards compatibility until the major version milestone. \\
    \midrule
    ${C}_{13}$ \textbf{User Interface and Outputs} \\ 
    \midrule
    \textbf{Definition}: This category refers to the user interface and outputs of the project. \\
    \textbf{Tags}: GUI and outputs, type:cli, end user \\
    \textbf{Example}: In scancode-toolkit issue \#389\footnote{\url{https://github.com/nexB/scancode-toolkit/issues/389}}, there is a CLI user interface issue where ScanCode's progress bar incorrectly displays each message on a new line instead of updating in place, likely caused by long filenames exceeding terminal width and disrupting the curses-based progress display synchronization. The problem affects the end user experience by making the command-line output messy and difficult to read during scans, requiring improvements to the GUI output formatting to properly handle terminal width constraints and filename truncation for better visual presentation. \\
    \midrule
    ${C}_{14}$ \textbf{Technical Debt} \\ 
    \midrule
    \textbf{Definition}: This category refers to addressing the technical debt of the project. \\
    \textbf{Tags}: technical-debt, technical debt \\
    \textbf{Example}: In tern issue \#212\footnote{\url{https://github.com/tern-tools/tern/issues/212}}, there is a technical debt issue where the \texttt{analyze\_docker\_image} function has grown overly complex with too many local variables, requiring significant code refactoring. The accumulated complexity is a result of deferred maintenance that should have been addressed earlier, necessitating a structural refactoring to separate key functionality into smaller, more manageable components or potentially creating a dedicated analysis module to improve code maintainability and reduce technical debt. \\ 
    \bottomrule
\end{longtable}

\begin{figure}[t]
    \centerline{\includegraphics[width=\linewidth]{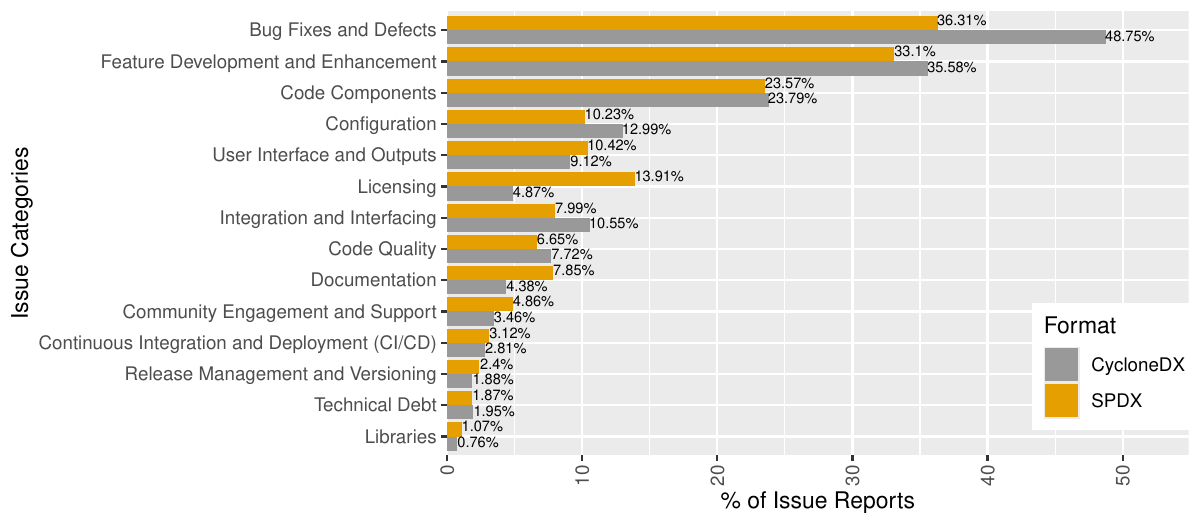}}
    \caption{The distribution of GitHub issue reports by category across \spdx and \cdx SBOM tools, as the share of categorized issue reports within each format that carry each category (an issue report can carry several categories).}
    \label{fig:tags_prevalence}
\end{figure}

\figref{fig:tags_prevalence} shows the share of issue reports that carry each category among the 26,772 categorized issue-report observations created since 2018 (14,763 for \cdx and 12,009 for \spdx, counted once per format), as classified by the LLM-assisted approach described in Section~\ref{sec:rq3-llm-classify}. An issue report can receive several categories, so the shares do not sum to $100\%$.

\textbf{Bug Fixes and Defects and Feature Development and Enhancement are the leading categories in both formats' tools in the LLM-assisted classification, but Bug Fixes and Defects weighs more heavily in \cdx.} According to the LLM-assisted classification in \figref{fig:tags_prevalence}, Bug Fixes and Defects accounts for $48.75\%$ of \cdx issue reports versus $36.31\%$ of \spdx issue reports, while Feature Development and Enhancement is similar ($35.58\%$ versus $33.10\%$) and Code Components is nearly identical ($23.79\%$ versus $23.57\%$). A cross-check that uses only the tag-based classification (shares of tagged issue reports, not shown in the figure) ranks Bug Fixes and Defects and Feature Development and Enhancement as the top two \cdx categories ($57.65\%$ and $40.91\%$) and also shows a larger Bug Fixes and Defects share in \cdx than in \spdx ($57.65\%$ versus $4.36\%$). For \spdx, however, Feature Development and Enhancement stays in the lead ($58.68\%$), yet Code Components is second ($18.78\%$), while Bug Fixes and Defects accounts for only $4.36\%$.

\textbf{Licensing and Documentation distinguish \spdx tools, whereas Configuration and Integration and Interfacing weigh more in \cdx tools.} Licensing accounts for $13.91\%$ of \spdx issue reports versus $4.87\%$ for \cdx, and Documentation for $7.85\%$ versus $4.38\%$, while Configuration ($12.99\%$ versus $10.23\%$) and Integration and Interfacing ($10.55\%$ versus $7.99\%$) weigh more in \cdx. The Licensing share in \spdx is $2.9\times$ that in \cdx. One hypothesis is that the \spdx licensing share reflects its origin as a licensing-focused format, and that the larger \cdx Bug Fixes and Defects share reflects its younger tools.

\begin{figure}[t]
    \centerline{\includegraphics[width=\linewidth]{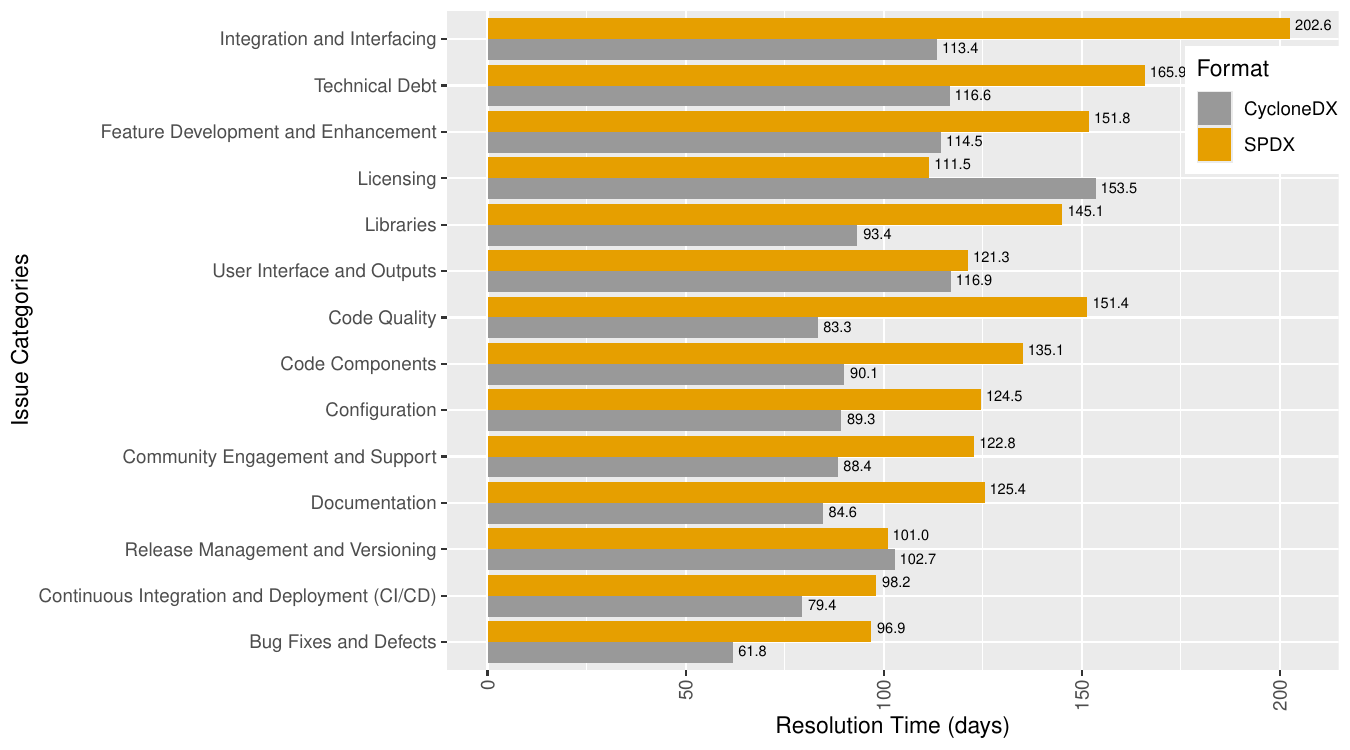}}
    \caption{Mean resolution time (in days) of closed issue reports by category for \spdx and \cdx tools.}
    \label{fig:tag_resolve_time}
\end{figure}

\textbf{\cdx tools resolve issue reports faster than \spdx tools in seven of the 14 categories, whereas \spdx tools are not significantly faster in any category.} To understand how each format handles different types of issue reports, we analyze the resolution time of each issue category within each format, as shown in \figref{fig:tag_resolve_time}. We compare the resolution times of the categories between the formats, employing the Mann-Whitney U test with Bonferroni correction over the 14 categories and Cliff's delta to assess effect size (\tabref{tab:category_resolution_time}). The mean resolution time of \cdx is $45.0\%$ lower than that of \spdx for Code Quality, $44.0\%$ lower for Integration and Interfacing, $36.2\%$ for Bug Fixes and Defects, $33.3\%$ for Code Components, $32.5\%$ for Documentation, $29.7\%$ for Technical Debt, and $24.6\%$ for Feature Development and Enhancement, with effect sizes that are negligible to small (largest: $|\delta| = 0.250$ for Code Quality). Licensing runs in the opposite direction (\cdx mean $153.5$ days versus \spdx $111.5$ days), but this difference is not significant after correction (adjusted $p=0.118$, nominal $p=0.008$), so we cannot conclude that \spdx resolves Licensing issue reports faster.

\begin{table}[t]
    \centering
    \caption{Mann-Whitney U test comparing the resolution time of closed issue reports per category between \cdx and \spdx tools (mean resolution time in days). The $p$-values are Bonferroni-corrected over the 14 categories. Cliff's delta and the effect size are shown for significant differences. Issue reports with several categories contribute to each of their categories.}
    \small
    \begin{tabular}{p{11em}rrrrl}
        \textbf{Category} & \textbf{\cdx} & \textbf{\spdx} & \textbf{Adj. $p$} & \textbf{$|\delta|$} & \textbf{Effect size}\\
        \midrule
        Bug Fixes and Defects & 61.8 & 96.9 & $<0.001$ & 0.125 & Negligible \\
        \midrule
        Code Quality & 83.3 & 151.4 & $<0.001$ & 0.250 & Small \\
        \midrule
        Code Components & 90.1 & 135.1 & $<0.001$ & 0.163 & Small \\
        \midrule
        Community Engagement and Support & 88.4 & 122.8 & 0.054 & - & - \\
        \midrule
        Configuration & 89.3 & 124.5 & 0.424 & - & - \\
        \midrule
        Continuous Integration and Deployment (CI/CD) & 79.4 & 98.2 & 1.000 & - & - \\
        \midrule
        Documentation & 84.6 & 125.4 & 0.046 & 0.103 & Negligible \\
        \midrule
        Feature Development and Enhancement & 114.5 & 151.8 & $<0.001$ & 0.132 & Negligible \\
        \midrule
        Integration and Interfacing & 113.4 & 202.6 & $<0.001$ & 0.150 & Small \\
        \midrule
        Libraries & 93.4 & 145.1 & 0.119 & - & - \\
        \midrule
        Licensing & 153.5 & 111.5 & 0.118 & - & - \\
        \midrule
        Release Management and Versioning & 102.7 & 101.0 & 1.000 & - & - \\
        \midrule
        Technical Debt & 116.6 & 165.9 & 0.041 & 0.173 & Small \\
        \midrule
        User Interface and Outputs & 116.9 & 121.3 & 1.000 & - & - \\
    \end{tabular}
    \label{tab:category_resolution_time}
\end{table}
 
\summarybox{RQ3: Most prevalent issue reports}{lightgray}{white}{Bug Fixes and Defects is the most prevalent category for both formats' tools and weighs more heavily in \cdx ($48.75\%$ versus $36.31\%$ of issue reports), while Licensing is a distinguishing category of \spdx ($13.91\%$ versus $4.87\%$). After Bonferroni correction, \cdx tools resolve issue reports faster than \spdx tools in seven of 14 categories (24.6\%--45.0\% lower mean resolution time, negligible to small effects), and \spdx tools are not significantly faster in any category. The category analysis includes 26,772 format-specific issue-report observations created since 2018 (99.7\% of the dataset). The LLM prompt was calibrated using a sample of 379 issue reports; the resulting classifications support aggregate-level comparisons.}
 \subsection{Prevalence of SBOM tools in OSS projects (RQ4)}

In this RQ, we investigate GitHub projects that use the \cdx or \spdx format, along with their associated tools, for their project's SBOM. By analyzing these OSS projects, we provide insights into the trends of the developer community and the characteristics of projects that use SBOM tools (\spdx or \cdx).
In addition, we analyze the prevalence of SBOM features in the OSS community. \spdx and \cdx tool developers may use this information to improve the features of their tools.

\subsubsection{Insights on the developer community of OSS projects that use SBOM tools.}

\begin{figure}[t]
    \centerline{\includegraphics[width=\linewidth]{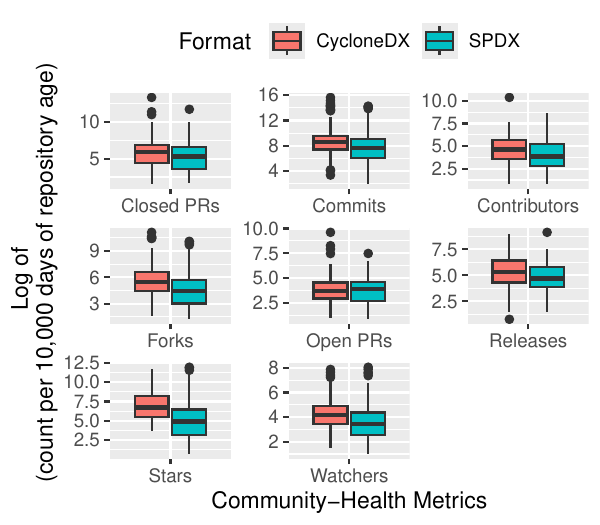}}
    \caption{Community-health metrics of top 250 projects that use \spdx tools versus top 250 projects that use \cdx tools. Counts are normalized by repository age (per 10,000 days) and shown on a log scale.}
    \label{fig:github_stats}
\end{figure} \begin{table*}
	\centering
	\caption{Mann-Whitney U test~\citep{mann1947test} applied to evaluate the difference between the community-health metrics of 250 projects that use \cdx tools and 250 projects that use \spdx tools. The $p$-values are Bonferroni-corrected over the ten metrics, and Cliff's delta is positive when \cdx values exceed \spdx values. $p < 0.05$ indicates a significant difference between the two distributions.}

	\begin{tabular}{lrrr}
		\toprule
		Metric & p-value & Cliff's delta &  Effect size \\
		\cmidrule(lrr){1-4}
			ClosedIssueReports & $1.95\times10^{-11}$ & $0.362$ & Medium\\
			ClosedPullRequest & $1.94\times10^{-07}$ & $0.290$ & Small \\
			Contributors & $2.69\times10^{-03}$ & $0.189$ & Small \\
			Forks & $2.69\times10^{-16}$ & $0.439$ & Medium\\
			OpenIssueReports & $9.32\times10^{-09}$ & $0.313$ & Small\\
		OpenPullRequest & $1.79\times10^{-05}$ & $0.241$ & Small\\
			Releases & $7.66\times10^{-09}$ & $0.308$ & Small\\
			Stars & $2.16\times10^{-25}$ & $0.552$ & Large\\
			Watchers & $2.49\times10^{-11}$ & $0.364$ & Medium\\
	        Commits & $4.20\times10^{-05}$ & $0.239$ & Small\\
		\bottomrule
	\end{tabular}

	\label{tab:projects_difference}
\end{table*}
 To understand the significance of projects that use the \spdx and \cdx tools, we assess the community health of identified projects that use the metrics of RQ2 (number of stars, watchers, forks, contributors, PRs, releases, and commits) plus the number of open and closed issue reports.
These metrics have previously been used in literature to assess many quality factors of software repositories, such as consistent documentation, number of downloads, popularity, and development activity~\cite{zerouali2019diversity}. These metrics also provide an idea about the health of the OSS community around these projects~\cite{chaoss}. 

\textbf{GitHub projects that use \cdx tools tend to be more active and mature than those that use \spdx tools, as reflected by higher counts of stars, watchers, forks, PRs, releases, contributors, and commits across key OSS community-health metrics.}
\figref{fig:github_stats} presents the values of the metrics of the projects that use \cdx or \spdx tools.
To compare the difference between projects that use \spdx tools and projects that use \cdx tools, we applied the Mann-Whitney U test~\citep{mann1947test} followed by the Bonferroni correction, on the non-normal distributions of the extracted metric values.
\tabref{tab:projects_difference} shows test results and the respective effect sizes\citep{effectSize}. It can be observed from this table that both sets are significantly different from each other for all the metrics after Bonferroni correction, with Cliff's delta effect sizes from small (0.19 for Contributors) to large (0.55 for Stars).

Projects that use \cdx tools have a higher median value (normalized by the age of the repository, per 10,000 days) in terms of stars (853 in \cdx versus only 61 in \spdx), forks (226 in \cdx versus 35 in \spdx), and watchers (62 in \cdx versus 28 in \spdx).
The higher numbers of stars and forks of projects that use \cdx tools can indicate higher trust and usage from the OSS community~\cite{borges2018s,whyFork}.
Furthermore, projects that use \cdx tools have a higher number of watchers, which could indicate
greater observability by the OSS community compared to projects that use \spdx tools~\cite{borges2018s}.

Projects that use \cdx tools exhibit significantly higher activity compared to those that use \spdx tools, as indicated by median values normalized by repository age (per 10,000 days). Specifically, these projects show higher median open PRs (26.1 in \cdx versus 2.4 in \spdx), median closed PRs (248 in \cdx versus 41 in \spdx), median releases (91 in \cdx versus 0 in \spdx), median contributors (99 in \cdx versus 46 in \spdx), and median commits (5,438 in \cdx versus 2,045 in \spdx).

These trends suggest that projects that use \cdx tools currently attract more attention and contributions from the OSS community. We hypothesize that this relates to the origins of the formats: \cdx was introduced as an SBOM format from the beginning, while \spdx initially focused only on licensing before expanding its scope, which may have influenced the types of projects each format attracted. Under this hypothesis, \cdx would see wider adoption in security-focused projects despite \spdx's head start (2010 versus 2017), while \spdx would remain more closely aligned with compliance-oriented projects, particularly those backed by the Linux Foundation~\citep{spdx_ref}.

\summarybox{RQ4: Community health of projects that use SBOM tools}{lightgray}{white}{Community-health metrics indicate notable differences between the top 250 projects that use \cdx tools and those that use \spdx tools. Projects that use \cdx tools tend to show higher levels of activity and engagement across various indicators, suggesting that SBOM format adoption may be associated with distinct community dynamics and maturity levels.}

\subsubsection{Insights on the language distribution of OSS projects that use SBOM tools.}

The choice of programming language for a project significantly impacts development efficiency, performance, maintainability \citep{10.1145/1985793.1985817}, security, and code quality \citep{10.1145/3126905}. To understand which programming-language-based projects prefer using which SBOM format (\cdx or \spdx), we classify the projects based on programming core, i.e., whether a project is written in a general-purpose language (GPL) or a domain-specific language (DSL).

\begin{figure}
    \centering

    \begin{subfigure}{\textwidth}
        \includegraphics[width=\linewidth]{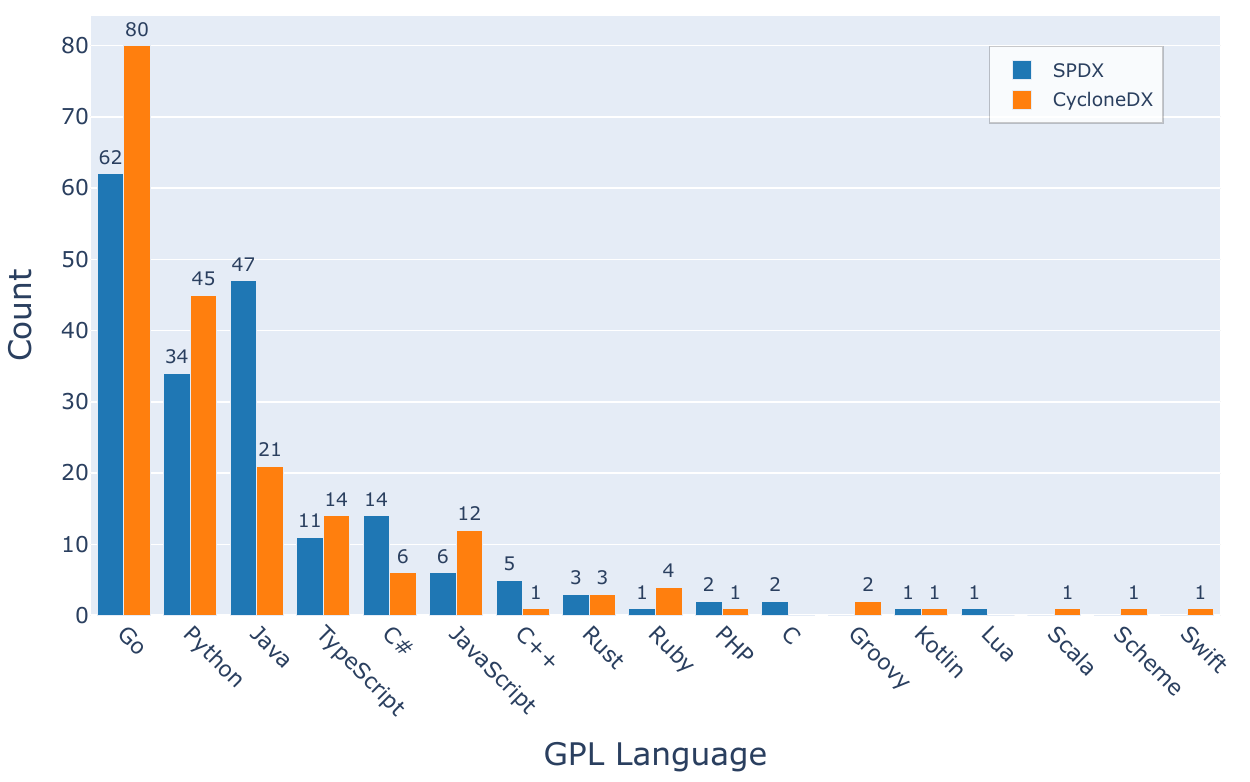}
    \end{subfigure}

    \begin{subfigure}{\textwidth}
        \includegraphics[width=\linewidth]{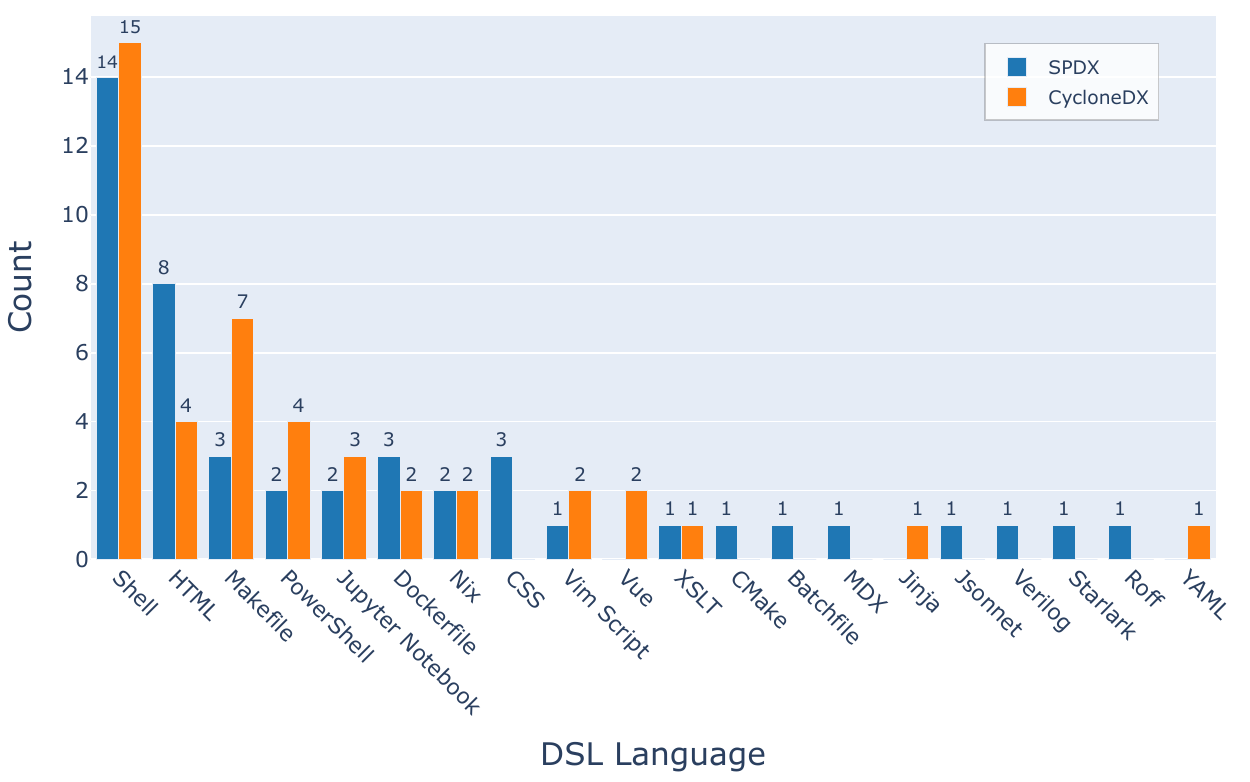}
    \end{subfigure}

    \caption{Programming languages of projects that use \cdx tools versus projects that use \spdx tools. A single project may use multiple languages.}
    \label{fig:language_stats}

\end{figure}
 
\figref{fig:language_stats} shows the language distribution of the projects that use GPL and DSL languages. We find that the usage of \cdx tools is more prevalent in Go and Python language based projects, whereas the usage of \spdx tools is more prevalent in Java and C\# language based projects. The association of Java and C\# language-based projects to \spdx usage aligns with \spdx's established position in enterprise and formal compliance environments. \spdx's comprehensive licensing and compliance focus corresponds to Java's enterprise heritage and C\#'s corporate ecosystem orientation. One conjecture is that the formal structure of \spdx documents suits the static typing and structured nature of these languages.

Conversely, Go-based projects demonstrate a marked preference for \cdx (80 projects compared to 62 for \spdx). This aligns with Go's prominence in containerized and microservices architectures, where \cdx's security vulnerability and supply chain analysis capabilities provide particular value. Python projects similarly favor \cdx tools (45 versus 34), which may relate to \cdx's approach to dependency tracking in dynamic development environments or to its emphasis on security-oriented metadata and integration with modern DevOps pipelines.

\summarybox{RQ4: Language Preference}{lightgray}{white}{We observe a language-based prevalence among projects that use different SBOM formats: the usage of \cdx tools is prevalent in projects written in Go and Python languages, while the usage of \spdx tools is prevalent in projects written in Java and C\# languages. While neither format is language-specific, the observed patterns may reflect the historical roots and typical adoption contexts of each format. We hypothesize that tool support is shaped by the leading languages among each format's users, which our data cannot confirm.}

\subsubsection{Tool preference across different SBOM projects.}

\textbf{The use cases supported by SBOM tools vary with project activity levels.}  
To examine whether more or less active projects tend to use tools associated with different use cases, we analyze project activity using two indicators: number of contributors and number of commits. First, we group projects into four quartiles based on contributor count and apply a chi-square test, finding a statistically significant difference among the quartiles ($p=3.39 \times 10^{-19}$), though with a small effect size ($\textit{Cramér's V}=0.093$).
Next, we group projects by commit count into quartiles and again observe a statistically significant difference ($p=5.56 \times 10^{-43}$), again with a small effect size ($\textit{Cramér's V}=0.127$)~\cite{cramer1999mathematical,effectSize}.

\begin{figure}
    \centering
    \includegraphics[width=\textwidth]{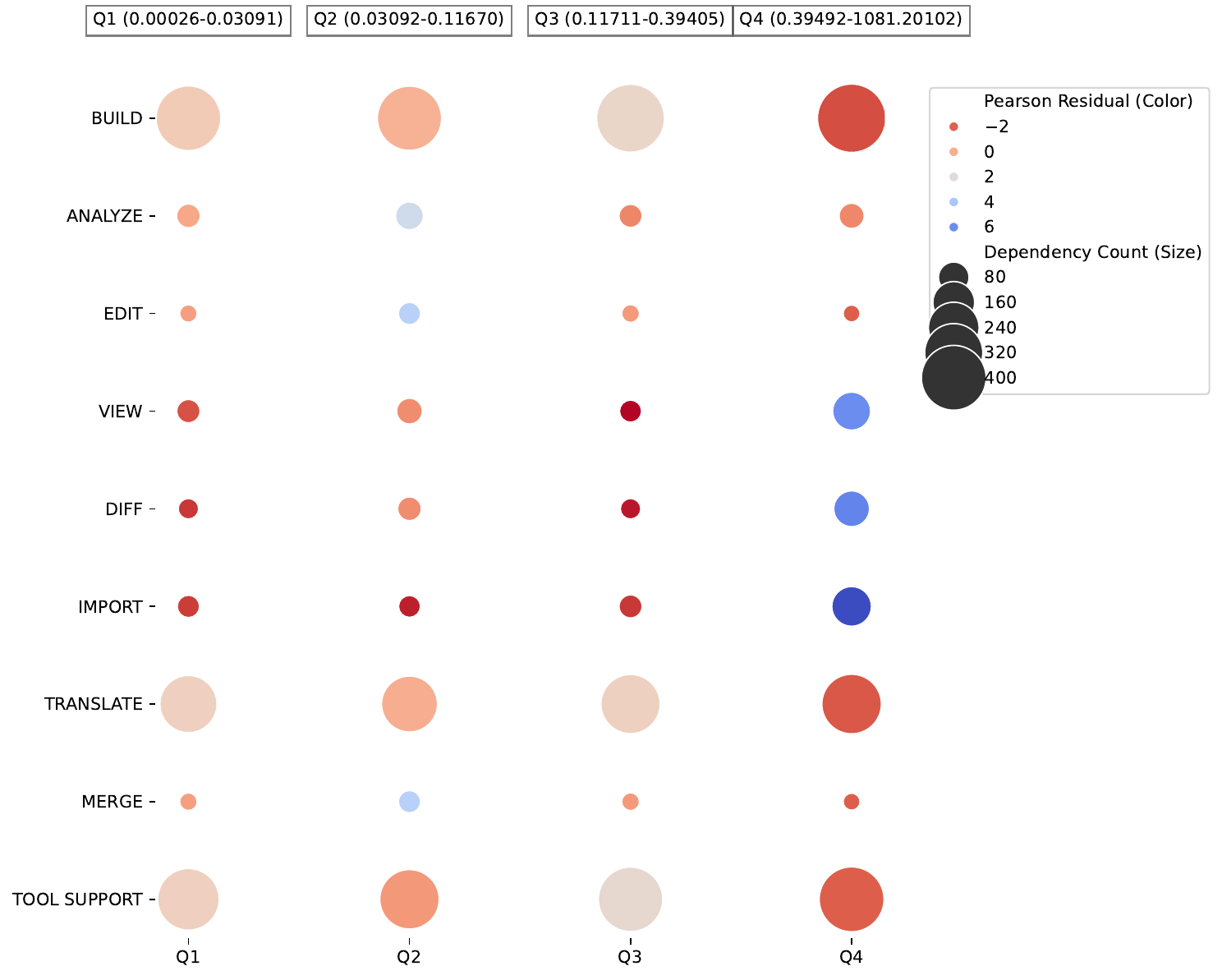}
    \caption{Visualization of the Pearson residuals to compare the Quartiles (by number of commits) with the use cases of the tools they use. The quartiles are normalized by the age of repository in days. Q1 = 0.00026--0.031, Q2 = 0.031--0.12, Q3 = 0.12--0.39, Q4 = 0.39--1081.20. The color of the bubble denotes the Pearson residual, and its size denotes the number of tool dependencies.}
    \label{fig:commit_quartiles}
\end{figure}

Figure~\ref{fig:commit_quartiles} presents the distribution of chi-square residuals across use cases and quartiles. In the figure, color reflects the Pearson residual of a use case within each activity level, while bubble size reflects the number of tool dependencies. These findings suggest that project activity is associated with differing preferences for SBOM tool use cases, potentially shaped by the project's scale, complexity, or workflow needs.

Our analysis reveals that the \emph{View}, \emph{Diff}, and \emph{Import} use cases from the \emph{Consume} category of the NTIA tool taxonomy exhibit the highest Pearson residuals in the top quartile (Q4), suggesting that projects with a higher number of commits are more likely to use tools supporting these use cases.
In contrast, the \emph{Analyze} and \emph{Edit} use cases from the \emph{Produce} category, and \emph{Merge} from the \emph{Transform} category are positively associated only with the second quartile (Q2), indicating that projects with comparatively few commits tend to use tools covering these specific functions.

\summarybox{RQ4: Tool usage across projects with varying activity levels}{lightgray}{white}{Projects with higher commit counts tend to use tools that support \emph{View}, \emph{Diff}, and \emph{Import} use cases, while projects in the second commit quartile more frequently use tools associated with \emph{Analyze}, \emph{Edit}, and \emph{Merge}. These patterns suggest that tool usage aligns with project activity levels, potentially reflecting different workflow requirements.}

  \section{Implications}
\label{sec:implications}

\paragraph{Implications for \spdx tool developers.}
The fact that \spdx-only tools are negatively associated with the \emph{Build} use case (RQ1, \tabref{tab:cdx_vs_spdx_residuals}), and that only 307 of the 1,394 single-format projects in RQ4 use \spdx tools, suggests that the \spdx tools we studied are less often invoked in automated build pipelines than \cdx tools. This might be partially explained by the fact that our search patterns cover only 11 \spdx tools versus 50 \cdx tools (Section~\ref{sec:method:rq4-validation}), or by the fact that platform-level \spdx generation, such as the export GitHub provides, is not captured by tool counts (RQ1). Furthermore, \cdx tools have lower mean issue-resolution times than \spdx tools in seven of 14 categories, by 24.6\%--45.0\%, with the largest gaps for Code Quality and Integration and Interfacing (RQ3, \tabref{tab:category_resolution_time}, negligible to small effects). These differences describe recorded resolution times only, i.e., further analysis is needed to understand why they arise. Finally, a scan of the \spdx tool repositories found that only 5 of 456 (1.1\%) advertise \spdx 3.0 support. This snapshot does not establish the reasons for the observed uptake or the challenges of migration.

\paragraph{Implications for \cdx tool developers.}
Bug Fixes and Defects accounts for $48.75\%$ of \cdx issue reports versus $36.31\%$ of \spdx issue reports (\figref{fig:tags_prevalence}). \cdx tools have a modestly lower recorded resolution time for these reports than \spdx tools, with a negligible effect size (Cliff's delta $= 0.125$, \tabref{tab:category_resolution_time}). The high prevalence of these reports motivates questions for future work: our taxonomy does not separate issue reports about the contents of generated SBOMs, so future studies could examine the nature of the bugs in both formats, whether they are comparable, and (e.g., through interviews with maintainers and users) whether defect-prevention practices would provide more value than further reducing resolution time. \cdx tools are also negatively associated with the \emph{Diff} use case (RQ1, \figref{fig:cdx_vs_spdx_residuals}), so the \cdx community should look into SBOM comparison features, for instance by adapting \spdx Diff tools to \cdx.

\paragraph{Implications for OSS SBOM tool developers.}
OSS tools are negatively associated with the \emph{Analyze} and \emph{View} use cases, with which proprietary tools are positively associated (RQ1, \tabref{tab:open_vs_prop_residuals} and \figref{fig:open_vs_prop_residuals}), so OSS developers of both formats should prioritize analytical features and SBOM visualization to close the gap with proprietary tools.

\paragraph{Implications for SBOM tool adopters.}
Adopters should consider the supporting tools when choosing a format, in addition to the specification itself. Concretely, (1) for SBOM \emph{generation} in a security-focused DevOps pipeline, \cdx tools show stronger \emph{Build} and \emph{Tool Support} coverage (RQ1) and are used more often in Go and Python projects (RQ4, \figref{fig:language_stats}). (2) For license compliance or SBOM \emph{translation}, \spdx tools show stronger \emph{Translate} and \emph{Diff} coverage (RQ1) and Licensing accounts for a larger share of \spdx issue reports ($13.91\%$ versus $4.87\%$, RQ3), consistent with its licensing heritage. Projects that use \spdx tools are more common in Java and C\# (RQ4). (3) For interoperability across both formats, dual-format tools are the most common choice, particularly among proprietary tools (64.52\%, RQ1).

Community-health metrics provide additional evidence about activity in the sampled repositories. The median normalized counts for \cdx tools exceed those for \spdx tools for the number of stars (46.9 versus 4.45), forks (20.8 versus 2.93), open and closed PRs (5.56 and 54 versus 0 and 0), and contributors (21.4 versus 7.29) (RQ2, \tabref{tab:metric_difference}). These metrics describe repository activity and engagement. While projects with higher commit counts more often use tools supporting \emph{View}, \emph{Diff}, and \emph{Import} (RQ4, \figref{fig:commit_quartiles}), this does not show that a more active tooling community produces better tools.

\paragraph{Implications for researchers.}
The sampled SBOM tool communities, represented by tools listed in the official \spdx and \cdx tool centers, their GitHub-hosted repositories, and identifiable contributing forks, are organized around competing formats. Within this dataset, their tool communities differ in use-case coverage (RQ1), community health (RQ2), issue report prevalence and resolution time (RQ3), and the profile of adopting projects (RQ4). A non-trivial share of the sampled tools supports multiple formats, and the formats continue to evolve, which enables cross-community comparisons that single-format settings do not allow. The number of closed \cdx issue reports begins to exceed the number of closed \spdx issue reports around the 2021 US executive order~\citep{eo14028} (\figref{fig:trend_of_issues}), which makes these communities a setting for studying how regulatory events and OSS dynamics interact, although our data cannot attribute the shift to the executive order.

RQ4 uses search patterns derived from Build-oriented tool invocations (i.e., commands that generate SBOMs) and therefore provides evidence primarily about the \emph{Build} use case (Section~\ref{sec:method:rq4-validation}). Understanding adoption of other NTIA categories, which may be used in-house or ad hoc, will require other data sources, perhaps interviews or focus groups with practitioners.
 \section{Threats to Validity}
\label{sec:threats}

\subsection{Construct Validity}

Deriving the RQ3 taxonomy from maintainer-assigned tags requires judgment since tags can be grouped differently, an issue report can span multiple categories, and a tag such as ``stale'' can have different meanings across repositories. We mitigate this threat through the tag-curation and author-review procedure in Section~\ref{sec:method:tag-grouping}. The first author grouped prevalent tags after removing uninformative tags and merging synonyms, then the last author independently reviewed the grouping before disagreements were resolved through discussion. Nonetheless, variation in maintainer tag use may introduce noise into the taxonomy.

To reduce dependence on inconsistent repository tags and to extend classification to untagged issue reports, an LLM assigns issue categories (Section~\ref{sec:rq3-llm-classify}) based on each issue report's title and a truncated body. These LLM-assisted category assignments are approximate classifications. Model and prompt choices may produce different category assignments. The classifier receives only issue titles and truncated bodies to keep inputs uniform and bounded, excluding discussion comments, linked PRs, and code context that maintainers could use to diagnose an issue report.

RQ4 statically identifies SBOM-generation patterns in repository files instead of dynamically configuring and executing the projects, so it does not establish that a matched invocation succeeds or produces a valid SBOM. A stratified random validation sample (Section~\ref{sec:method:rq4-validation}) found that 81 of the 87 retrievable entries (93.1\%) were genuine Build-oriented SBOM-tool invocations. The remaining 6 comprised 4 unrelated matches (false positives) and 2 indirect references. The sample and the validation label for each of the 87 retrievable entries are in the replication package.

\spdx and \cdx support several serializations, of which JSON and XML are common to both. We manually compared their specifications and found that the JSON and XML forms cover the same SBOM documentation features relevant to RQ1. We therefore use the XML specifications for this comparison of SBOM documentation features; the tool and repository analyses are not restricted to XML-based tools. Unexamined differences in how tools implement these serializations may nevertheless affect the results, and future work should compare them systematically.

\subsection{External Validity}

RQ3 analyzes issue reports in GitHub repositories of \cdx and \spdx tools. Its findings may therefore not generalize to other SBOM formats, hosting platforms, or discussions outside GitHub. We focus on \spdx and \cdx since they are the two leading formats with the most extensive tool support~\citep{xia2023empirical,mirakhorli2024landscape}.

Our tool dataset covers the tools listed in the official \cdx and \spdx tool centers, their GitHub-hosted repositories, and contributing forks identifiable within the data collection window. Our tool-level findings are therefore limited to that population. They do not cover commercial or closed-source tools absent from the tool centers or tools hosted outside GitHub. As a coverage cross-check, we compared the dataset with the GitHub repositories linked in the independently maintained \textit{awesome-sbom} catalog~\citep{awesomesbom}. Fourteen of its 29 linked GitHub repositories (48.3\%) are represented in our dataset, counting 12 exact matches and two repository-name matches. These include \texttt{anchore/syft}, \texttt{microsoft/sbom-tool}, and \texttt{CycloneDX/cdxgen}. This descriptive cross-check does not establish coverage of tools beyond the official centers. Future work should incorporate additional tool data sources.

Since the RQ4 queries were derived from Build-oriented invocations, RQ4 provides evidence primarily about the \emph{Build} use case and cannot show that Build is leading outside this sampling frame (Section~\ref{sec:method:rq4-validation}). Our GitHub-based analysis also does not measure industrial adoption or use of \spdx and \cdx.

The contributing forks account for the larger number of \spdx tool repositories (Section~\ref{sec:method:tools-extension}) and may lower the \spdx medians in RQ2, and we did not repeat the comparison without them. RQ4 compares the top 250 of 307 projects that use \spdx tools (81.4\%) with the top 250 of 1,087 projects that use \cdx tools (23.0\%), so selecting by stars is stricter for \cdx and favors it on stars and related popularity metrics.

Our comparison is contextualized by \spdx 2.3~\citep{spdx_about} and \cdx 1.5~\citep{CycloneDXSpecs}, the latest specifications available at data collection, while the tools and issue reports we study span several specification releases. Our findings may not generalize to tools that adopt \spdx 3.0~\citep{spdx30release}, which may substantially change the \spdx tooling landscape. Of the 456 \spdx tool repositories, nine matched \spdx 3.0 keywords, four of which are specification, model, or meeting repositories rather than tools, so only 5 (1.1\%) explicitly advertised \spdx 3.0 tool support at the time of our analysis. The remaining repositories did not explicitly advertise such support, which does not establish whether they had migrated. Since \spdx 3.0 implements a JSON-LD format and does not support the \spdx 2.x Tag/Value, YAML, RDF/XML, and spreadsheet serializations~\citep{spdx30release,spdx30diffs}, tools require active updates to support it. Our findings therefore provide a baseline for evaluating post-3.0 changes, but tool availability and issue-category distributions may change over time.

\subsection{Internal Validity}

Repeated LLM runs occasionally assign different categories to the same issue report, even at temperature 0, and an issue report can receive up to four categories, so category shares do not partition the dataset. These limitations affect how strongly the classifications support aggregate \spdx--\cdx comparisons and category-share estimates.

To limit these threats, we used the same model, final prompt, and temperature for the primary classification run and use the labels only for aggregate \spdx--\cdx comparisons. Explanations for these patterns remain hypotheses that require validation through developer interviews or surveys.
 \section{Conclusion}

This study compares the tools of the two leading SBOM formats, \spdx and \cdx, as listed in the official tool centers and hosted on GitHub, across their OSS and proprietary tools, the community health of those tools, their issue reports, and the projects that use them. We find that proprietary tools support a broader range of use cases than OSS tools, and that \cdx tools show higher developer engagement and faster issue resolution than \spdx tools, with the differences varying in size. For practitioners and policymakers, these findings suggest considering the maturity and vibrancy of the supporting tools, in addition to the specification itself, when choosing an SBOM format. For tool developers, they point to areas for improvement in functionality and community support.
 \section*{Acknowledgement}

We acknowledge Ankit Hans and Ujjwal Agarwal for their assistance in the early stages of idea formulation and data collection. We also sincerely thank Ms.~Kate Stewart\footnote{\url{https://www.rsaconference.com/experts/kate-stewart}}, VP for Dependable Embedded Systems at the Linux Foundation, and Mr.~Gary O'Neall\footnote{\url{https://github.com/goneall}}, co-lead of the SPDX technical workgroup, both key contributors from the Linux Foundation to SPDX, for their invaluable insights and feedback during our investigation. \section*{Declarations}

\subsection*{Funding}
This research was supported by the NSERC Discovery Grant.

\subsection*{Conflicts of interest/Competing interests}
The authors have no competing interests to declare that are relevant to the content of this article.

\subsection*{Ethics approval}
This research does not involve human participants or animals as study subjects.

\subsection*{Consent to participate}
No human subjects were involved in this study.

\subsection*{Data availability}
The datasets generated and analyzed during this study are available in the replication package~\citep{replication}.
 \bibliographystyle{spbasic}
\bibliography{References}

\begin{thebibliography}{62}
\providecommand{\natexlab}[1]{#1}
\providecommand{\url}[1]{{#1}}
\providecommand{\urlprefix}{URL }
\expandafter\ifx\csname urlstyle\endcsname\relax
  \providecommand{\doi}[1]{DOI~\discretionary{}{}{}#1}\else
  \providecommand{\doi}{DOI~\discretionary{}{}{}\begingroup \urlstyle{rm}\Url}\fi
\providecommand{\eprint}[2][]{\url{#2}}

\bibitem[{{awesomeSBOM}(2024)}]{awesomesbom}
{awesomeSBOM} (2024) awesome-sbom: A curated list of {SBOM} (software bill of materials) related tools, frameworks, blogs, podcasts, and resources. \url{https://github.com/awesomeSBOM/awesome-sbom}

\bibitem[{Berger(2023)}]{dynatraceWhatLog4Shell}
Berger A (2023) {W}hat is {L}og4{S}hell? {T}he {L}og4j vulnerability explained (and what to do about it): dynatrace.com. \url{https://www.dynatrace.com/news/blog/what-is-log4shell/}, log4j vulnerability explained

\bibitem[{Bhattacharya and Neamtiu(2011)}]{10.1145/1985793.1985817}
Bhattacharya P, Neamtiu I (2011) Assessing programming language impact on development and maintenance: A study on c and c++. In: Proceedings of the 33rd International Conference on Software Engineering, Association for Computing Machinery, New York, NY, USA, ICSE '11, p 171–180, \doi{10.1145/1985793.1985817}, \urlprefix\url{https://doi.org/10.1145/1985793.1985817}

\bibitem[{Bi et~al.(2024)Bi, Xia, Xing, Lu, and Zhu}]{bi2023way}
Bi T, Xia B, Xing Z, Lu Q, Zhu L (2024) On the way to {SBOMs}: Investigating design issues and solutions in practice. ACM Transactions on Software Engineering and Methodology 33(6):1--25, \doi{10.1145/3654442}

\bibitem[{Borges and Valente(2018)}]{borges2018s}
Borges H, Valente MT (2018) What’s in a github star? understanding repository starring practices in a social coding platform. Journal of Systems and Software 146:112--129

\bibitem[{Campbell et~al.(2013)Campbell, Quincy, Osserman, and Pedersen}]{campbell2013coding}
Campbell JL, Quincy C, Osserman J, Pedersen OK (2013) Coding in-depth semistructured interviews: Problems of unitization and intercoder reliability and agreement. Sociological methods \& research 42(3):294--320

\bibitem[{{CHAOSS Project}(2024)}]{chaoss}
{CHAOSS Project} (2024) Community health analytics in open source software ({CHAOSS}): Metrics and metrics models. \url{https://chaoss.community/kb-metrics-and-metrics-models/}

\bibitem[{Chopra et~al.(2021)Chopra, Mo, Dodson, Beschastnikh, Fels, and Yoon}]{chopra2021alex}
Chopra A, Mo M, Dodson S, Beschastnikh I, Fels SS, Yoon D (2021) "@ alex, this fixes\# 9": Analysis of referencing patterns in pull request discussions. Proceedings of the ACM on human-computer interaction 5(CSCW2):1--25

\bibitem[{Cliff(1993)}]{cliff1993dominance}
Cliff N (1993) Dominance statistics: Ordinal analyses to answer ordinal questions. Psychological Bulletin 114(3):494--509, \doi{10.1037/0033-2909.114.3.494}

\bibitem[{Cochran(1977)}]{cochran1977sampling}
Cochran WG (1977) Sampling Techniques, 3rd edn. John Wiley \& Sons, New York

\bibitem[{Cohen(1992)}]{effectSize}
Cohen J (1992) A power primer. Psychological Bulletin 112(1):155--159, \doi{10.1037/0033-2909.112.1.155}

\bibitem[{Cox(2019)}]{cox2019surviving}
Cox R (2019) Surviving software dependencies. Communications of the ACM 62(9):36--43, \doi{10.1145/3347446}

\bibitem[{Cram{\'e}r(1999)}]{cramer1999mathematical}
Cram{\'e}r H (1999) Mathematical methods of statistics, vol~43. Princeton university press

\bibitem[{{CycloneDX Project}(2017)}]{cyclonedxspec}
{CycloneDX Project} (2017) {CycloneDX} specification repository. \url{https://github.com/CycloneDX/specification}, initial commit on May 29, 2017. CycloneDX 1.6 released on April 9, 2024

\bibitem[{{CycloneDX Project}(2025)}]{cyclonedx_ref}
{CycloneDX Project} (2025) {CycloneDX} history. \url{https://cyclonedx.org/about/history/}

\bibitem[{Das(2025)}]{ravi}
Das R (2025) {SBOM} formats explained: Guide for enterprises. TechTarget, \url{https://www.techtarget.com/cybersecurity/tip/SBOM-formats-explained-Guide-for-enterprises}

\bibitem[{Diglio(2021)}]{diglio2021spdx}
Diglio A (2021) Generating software bills of materials ({SBOMs}) with {SPDX} at {Microsoft}. Engineering@Microsoft, \url{https://devblogs.microsoft.com/engineering-at-microsoft/generating-software-bills-of-materials-sboms-with-spdx-at-microsoft/}

\bibitem[{Efron(1979)}]{efron1979bootstrap}
Efron B (1979) Bootstrap methods: Another look at the jackknife. The Annals of Statistics 7(1):1--26, \doi{10.1214/aos/1176344552}

\bibitem[{{European Parliament and Council of the European Union}(2024)}]{cra2024}
{European Parliament and Council of the European Union} (2024) Regulation ({EU}) 2024/2847 on horizontal cybersecurity requirements for products with digital elements ({Cyber Resilience Act}). Official Journal of the European Union, \url{https://eur-lex.europa.eu/eli/reg/2024/2847/oj}

\bibitem[{Foundjem et~al.(2021)Foundjem, Eghan, and Adams}]{foundjem2021onboarding}
Foundjem A, Eghan E, Adams B (2021) Onboarding vs. diversity, productivity and quality: empirical study of the openstack ecosystem. In: 2021 IEEE/ACM 43rd International Conference on Software Engineering (ICSE), IEEE, pp 1033--1045

\bibitem[{Gao et~al.(2024)Gao, He, Xie et~al.}]{gao2024characterizing}
Gao K, He R, Xie B, et~al. (2024) Characterizing deep learning package supply chains in {PyPI}: Domains, clusters, and disengagement. ACM Transactions on Software Engineering and Methodology 33(4):1--27

\bibitem[{{GitHub}(2026)}]{githubsbomexport}
{GitHub} (2026) Exporting a software bill of materials for your repository. GitHub Docs, \url{https://docs.github.com/en/code-security/how-tos/secure-your-supply-chain/establish-provenance-and-integrity/exporting-a-software-bill-of-materials-for-your-repository}

\bibitem[{Gkortzis et~al.(2021)Gkortzis, Feitosa, and Spinellis}]{gkortzis2021software}
Gkortzis A, Feitosa D, Spinellis D (2021) Software reuse cuts both ways: An empirical analysis of its relationship with security vulnerabilities. Journal of Systems and Software 172:110653

\bibitem[{Hendrick(2022)}]{linuxReadiness}
Hendrick S (2022) The state of software bill of materials ({SBOM}) and cybersecurity readiness. Linux Foundation Research, \url{https://www.linuxfoundation.org/research/the-state-of-software-bill-of-materials-sbom-and-cybersecurity-readiness-gated}, linux Foundation

\bibitem[{Herraiz et~al.(2011)Herraiz, Shihab, Nguyen, and Hassan}]{6079845}
Herraiz I, Shihab E, Nguyen TH, Hassan AE (2011) Impact of installation counts on perceived quality: A case study on debian. In: 2011 18th Working Conference on Reverse Engineering, pp 219--228, \doi{10.1109/WCRE.2011.34}

\bibitem[{{ISO/IEC}(2015)}]{swid_ref}
{ISO/IEC} (2015) {ISO/IEC} 19770-2:2015 information technology --- software asset management --- part 2: Software identification tag. \url{https://www.iso.org/standard/65666.html}

\bibitem[{Jiang et~al.(2017)Jiang, Lo, He, Xia, Kochhar, and Zhang}]{whyFork}
Jiang J, Lo D, He J, Xia X, Kochhar PS, Zhang L (2017) Why and how developers fork what from whom in github. Empirical Software Engineering 22, \doi{10.1007/s10664-016-9436-6}

\bibitem[{Jiao et~al.(2000)Jiao, Tseng, Ma, and Zou}]{boms}
Jiao J, Tseng MM, Ma Q, Zou Y (2000) Generic bill-of-materials-and-operations for high-variety production management. Concurrent Engineering: Research and Applications 8(4):297--321, \doi{10.1177/1063293X0000800404}

\bibitem[{Kikas et~al.(2015)Kikas, Dumas, and Pfahl}]{kikas2015issue}
Kikas R, Dumas M, Pfahl D (2015) Issue dynamics in github projects. In: Product-Focused Software Process Improvement: 16th International Conference, PROFES 2015, Bolzano, Italy, December 2-4, 2015, Proceedings 16, Springer, pp 295--310

\bibitem[{Krippendorff(2019)}]{krippendorff2019content}
Krippendorff K (2019) Content Analysis: An Introduction to Its Methodology, 4th edn. SAGE Publications, \doi{10.4135/9781071878781}

\bibitem[{Ladisa et~al.(2023)Ladisa, Plate, Martinez, and Barais}]{ladisa2022taxonomy}
Ladisa P, Plate H, Martinez M, Barais O (2023) {SoK}: Taxonomy of attacks on open-source software supply chains. In: 2023 IEEE Symposium on Security and Privacy (SP), IEEE, pp 1509--1526, \doi{10.1109/SP46215.2023.10179304}

\bibitem[{Landis and Koch(1977)}]{landis1977measurement}
Landis JR, Koch GG (1977) The measurement of observer agreement for categorical data. Biometrics 33(1):159--174, \doi{10.2307/2529310}

\bibitem[{Mann and Whitney(1947)}]{mann1947test}
Mann HB, Whitney DR (1947) On a test of whether one of two random variables is stochastically larger than the other. The Annals of Mathematical Statistics 18(1):50--60, \doi{10.1214/aoms/1177730491}

\bibitem[{Martin(2020)}]{9174365}
Martin RA (2020) Visibility \& control: Addressing supply chain challenges to trustworthy software-enabled things. In: 2020 IEEE Systems Security Symposium (SSS), pp 1--4, \doi{10.1109/SSS47320.2020.9174365}

\bibitem[{Mens et~al.(2017)Mens, Adams, and Marsan}]{mens2017towards}
Mens T, Adams B, Marsan J (2017) Towards an interdisciplinary, socio-technical analysis of software ecosystem health. arXiv preprint arXiv:171104532

\bibitem[{Mirakhorli et~al.(2024)Mirakhorli, Garcia, Dillon, Laporte, Morrison, Lu, Koscinski, and Enoch}]{mirakhorli2024landscape}
Mirakhorli M, Garcia D, Dillon S, Laporte K, Morrison M, Lu H, Koscinski V, Enoch C (2024) A landscape study of open source and proprietary tools for software bill of materials (sbom). arXiv preprint arXiv:240211151

\bibitem[{Muir{\'\i}(2019)}]{muiri2019framing}
Muir{\'\i} {\'E}{\'O} (2019) Framing software component transparency: Establishing a common software bill of material ({SBOM}). NTIA Multistakeholder Process on Software Component Transparency, \url{https://www.ntia.gov/files/ntia/publications/framingsbom_20191112.pdf}, national Telecommunications and Information Administration

\bibitem[{Nocera et~al.(2023)Nocera, Romano, Di~Penta, Francese, and Scanniello}]{nocera2023software}
Nocera S, Romano S, Di~Penta M, Francese R, Scanniello G (2023) Software bill of materials adoption: A mining study from github. In: 2023 IEEE International Conference on Software Maintenance and Evolution (ICSME), IEEE, pp 39--49

\bibitem[{NTIA(2021)}]{sbommeeting}
NTIA (2021) Ntia software component transparency. \url{https://www.ntia.gov/other-publication/ntia-software-component-transparency}, [SBOM Virtual Multistakeholder Meeting]

\bibitem[{{OpenAI}(2023)}]{openai2023outage}
{OpenAI} (2023) March 20 {ChatGPT} outage: Here's what happened. \url{https://openai.com/index/march-20-chatgpt-outage/}

\bibitem[{OWASP(2023{\natexlab{a}})}]{CycloneDXSpecs}
OWASP (2023{\natexlab{a}}) Official cyclonedx specification documentation json reference v1.5. \url{https://cyclonedx.org/docs/1.5/json/}, [CycloneDx Documentation]

\bibitem[{OWASP(2023{\natexlab{b}})}]{CycloneDXTools}
OWASP (2023{\natexlab{b}}) Official cyclonedx tool center. \url{https://cyclonedx.org/tool-center/}, [CycloneDx Tools]

\bibitem[{{President of the United States}(2021)}]{eo14028}
{President of the United States} (2021) Executive order 14028: Improving the nation's cybersecurity. Federal Register 86(93):26633--26647, \url{https://www.federalregister.gov/documents/2021/05/17/2021-10460/improving-the-nations-cybersecurity}

\bibitem[{Ray et~al.(2017)Ray, Posnett, Devanbu, and Filkov}]{10.1145/3126905}
Ray B, Posnett D, Devanbu P, Filkov V (2017) A large-scale study of programming languages and code quality in github. Commun ACM 60(10):91–100, \doi{10.1145/3126905}, \urlprefix\url{https://doi.org/10.1145/3126905}

\bibitem[{Romano et~al.(2006)Romano, Kromrey, Coraggio, and Skowronek}]{romano2006appropriate}
Romano J, Kromrey JD, Coraggio J, Skowronek J (2006) Appropriate statistics for ordinal level data: Should we really be using t-test and {Cohen's} d for evaluating group differences on the {NSSE} and other surveys? In: Annual Meeting of the Florida Association of Institutional Research

\bibitem[{{SolarWinds Corporation}(2020)}]{solarwinds2020form8k}
{SolarWinds Corporation} (2020) Form 8-k, current report. U.S. Securities and Exchange Commission, \url{https://www.sec.gov/Archives/edgar/data/1739942/000162828020017451/swi-20201214.htm}

\bibitem[{{SPDX}(2023{\natexlab{a}})}]{spdx_tools_opesource}
{SPDX} (2023{\natexlab{a}}) {Official SPDX Open Source Tools}. \url{https://spdx.dev/spdx-tools}

\bibitem[{{SPDX}(2023{\natexlab{b}})}]{spdx_about}
{SPDX} (2023{\natexlab{b}}) {Official SPDX Specification Documentation v2.3}. \url{https://spdx.github.io/spdx-spec/v2.3/package-information/}, [SPDX Documentation]

\bibitem[{{SPDX Project}(2024{\natexlab{a}})}]{spdx30diffs}
{SPDX Project} (2024{\natexlab{a}}) Differences from previous editions: Using {SPDX}. \url{https://spdx.github.io/using/diffs-from-previous-editions/}

\bibitem[{{SPDX Project}(2024{\natexlab{b}})}]{spdx_ref}
{SPDX Project} (2024{\natexlab{b}}) {SPDX}: About. \url{https://spdx.dev/about/}

\bibitem[{{SPDX Project}(2024{\natexlab{c}})}]{spdx30release}
{SPDX Project} (2024{\natexlab{c}}) {SPDX} specification v3.0 release notes. \url{https://github.com/spdx/spdx-spec/releases/tag/v3.0}

\bibitem[{van~der Storm(2007)}]{van2007component}
van~der Storm T (2007) Component-based configuration, integration and delivery. PhD thesis, Universiteit van Amsterdam

\bibitem[{Tan et~al.(2020)Tan, Zhou, and Sun}]{tan2020first}
Tan X, Zhou M, Sun Z (2020) A first look at good first issues on {GitHub}. In: Proceedings of the 28th ACM Joint Meeting on European Software Engineering Conference and Symposium on the Foundations of Software Engineering, pp 398--409

\bibitem[{Telecommunications and (NTIA)(2021{\natexlab{a}})}]{NTIA2021}
Telecommunications N, (NTIA) IA (2021{\natexlab{a}}) {Minimum Elements for a Software Bill of Materials (SBOM)}. \urlprefix\url{https://www.ntia.doc.gov/files/ntia/publications/sbom_minimum_elements_report.pdf}, [NTIA2021]

\bibitem[{Telecommunications and (NTIA)(2021{\natexlab{b}})}]{NTIATaxonomy}
Telecommunications N, (NTIA) IA (2021{\natexlab{b}}) Sbom tool classification taxonomy. \urlprefix\url{https://www.ntia.gov/files/ntia/publications/ntia_sbom_tooling_taxonomy-2021mar30.pdf}, [NTIATaxonomy]

\bibitem[{Waltermire et~al.(2016)Waltermire, Feldman, and Witte}]{waltermire2016improving}
Waltermire DA, Feldman L, Witte GA (2016) Improving security and software management through the use of {SWID} tags. NIST ITL Bulletin, \url{https://csrc.nist.gov/pubs/itlb/2016/07/improving-security-software-management-using-swid/final}, national Institute of Standards and Technology

\bibitem[{Worthington(2023)}]{forrester}
Worthington J (2023) The world lags with {SBOM} requirements, but likely not for long. Forrester Blog, \url{https://www.forrester.com/blogs/the-world-lags-with-sbom-requirements-but-likely-not-for-long/}, forrester Research

\bibitem[{Xia et~al.(2023)Xia, Bi, Xing, Lu, and Zhu}]{xia2023empirical}
Xia B, Bi T, Xing Z, Lu Q, Zhu L (2023) An empirical study on software bill of materials: Where we stand and the road ahead. In: 2023 IEEE/ACM 45th International Conference on Software Engineering (ICSE), IEEE, pp 2630--2642

\bibitem[{Xia et~al.(2024)Xia, Zhang, Liu, Lu, Xing, and Zhu}]{xia2023trust}
Xia B, Zhang D, Liu Y, Lu Q, Xing Z, Zhu L (2024) Trust in software supply chains: Blockchain-enabled {SBOM} and the {AIBOM} future. In: Proceedings of the 2024 ACM/IEEE 4th International Workshop on Engineering and Cybersecurity of Critical Systems (EnCyCriS) and 2024 IEEE/ACM Second International Workshop on Software Vulnerability, ACM, pp 12--19, \doi{10.1145/3643662.3643957}

\bibitem[{Zerouali et~al.(2019)Zerouali, Mens, Robles, and Gonzalez-Barahona}]{zerouali2019diversity}
Zerouali A, Mens T, Robles G, Gonzalez-Barahona JM (2019) On the diversity of software package popularity metrics: An empirical study of npm. In: 2019 IEEE 26th international conference on software analysis, Evolution and Reengineering (SANER), IEEE, pp 589--593

\bibitem[{Zhao et~al.(2026)Zhao, Bangash, Ge, Singh, Wang, and Adams}]{replication}
Zhao Z, Bangash AA, Ge T, Singh A, Wang Z, Adams B (2026) Replication package for ``the state of the {SBOM} tool ecosystems: A comparative analysis of {SPDX} and {CycloneDX}''. \url{https://github.com/SAILResearch/26-ali-sbom-emse}

\bibitem[{Zimmermann(2016)}]{zimmermann2016card}
Zimmermann T (2016) Card-sorting: From text to themes. In: Perspectives on data science for software engineering, Elsevier, pp 137--141

\end{thebibliography}

\appendix
\section{Appendix}
\label{sec:appendix}

\setlength\LTleft{0pt}
\setlength\LTright{0pt}

\begin{longtable}{@{\extracolsep{\fill}}p{10em}cccccccccccc}

    \caption{Manual classification of 170 SBOM tools according to the NTIA tool taxonomy.} \label{tab:tool_classification}                                                                                                                                                                                              \\
    \toprule
    \multicolumn{3}{c}{}                                & \multicolumn{9}{c}{\bf NTIA Tool Taxonomy use case}                                                                                                                                                                                                             \\
    \cmidrule(lr){4-12}
    \multicolumn{1}{c}{}                                & \multicolumn{1}{c}{}                                & \multicolumn{1}{c}{}                        & \multicolumn{3}{c}{\bf Produce} & \multicolumn{3}{c}{\bf Consume} & \multicolumn{3}{c}{\bf Transform}                                                       \\
    \cmidrule(lr){4-6} \cmidrule(lr){7-9} \cmidrule(lr){10-12}
    \bf Name                                            & \rotatebox[origin=c]{90}{OSS}                & \rotatebox[origin=c]{90}{\makecell{Dual-                                                                                                                                                               \\ format}}  & \rotatebox[origin=c]{90}{Build} & \rotatebox[origin=c]{90}{Analyze} & \rotatebox[origin=c]{90}{Edit} & \rotatebox[origin=c]{90}{View} & \rotatebox[origin=c]{90}{Diff} & \rotatebox[origin=c]{90}{Import} & \rotatebox[origin=c]{90}{Translate} & \rotatebox[origin=c]{90}{Merge} & \rotatebox[origin=c]{90}{Tool Support} \\
    \midrule
    \endfirsthead

    \multicolumn{12}{c}
    {{\bfseries continued from previous page}}                                                                                                                                                                                                                                                                            \\
    \toprule
    \multicolumn{3}{c}{}                                & \multicolumn{9}{c}{\bf NTIA Tool Taxonomy use case}                                                                                                                                                                                                                    \\
    \cmidrule(lr){4-12}
    \multicolumn{1}{c}{}                                & \multicolumn{1}{c}{}                                & \multicolumn{1}{c}{}                        & \multicolumn{3}{c}{\bf Produce} & \multicolumn{3}{c}{\bf Consume} & \multicolumn{3}{c}{\bf Transform}                                                       \\
    \cmidrule(lr){4-6} \cmidrule(lr){7-9} \cmidrule(lr){10-12}
    \bf Name                                            & \rotatebox[origin=c]{90}{OSS}                & \rotatebox[origin=c]{90}{\makecell{Dual-                                                                                                                                                               \\ format}}  & \rotatebox[origin=c]{90}{Build} & \rotatebox[origin=c]{90}{Analyze} & \rotatebox[origin=c]{90}{Edit} & \rotatebox[origin=c]{90}{View} & \rotatebox[origin=c]{90}{Diff} & \rotatebox[origin=c]{90}{Import} & \rotatebox[origin=c]{90}{Translate} & \rotatebox[origin=c]{90}{Merge} & \rotatebox[origin=c]{90}{Tool Support} \\
    \midrule
    \endhead

    \endfoot

    \hline \hline
    \endlastfoot

    action-owasp-dependecy-track-check                  & \cmark                                              & \xmark                                      & \cmark                          & \cmark                          & \xmark                            & \cmark & \xmark & \xmark & \xmark & \xmark & \xmark \\
    apt2sbom                                            & \cmark                                              & \cmark                                      & \cmark                          & \xmark                          & \xmark                            & \xmark & \xmark & \xmark & \xmark & \xmark & \xmark \\
    Auditjs                                             & \cmark                                              & \xmark                                      & \cmark                          & \cmark                          & \xmark                            & \xmark & \xmark & \xmark & \xmark & \xmark & \cmark \\
    Augur                                               & \cmark                                              & \xmark                                      & \cmark                          & \cmark                          & \xmark                            & \xmark & \xmark & \xmark & \xmark & \xmark & \xmark \\
    bom                                                 & \cmark                                              & \xmark                                      & \cmark                          & \cmark                          & \xmark                            & \cmark & \xmark & \xmark & \xmark & \xmark & \cmark \\
    Cavil                                               & \cmark                                              & \xmark                                      & \xmark                          & \xmark                          & \xmark                            & \cmark & \xmark & \xmark & \xmark & \xmark & \xmark \\
\end{longtable} 

\begin{table}[t]
	\caption{\spdx and \cdx Tools, and their corresponding search patterns. We use these search patterns to identify GitHub projects that leverage these tools.}
	\begin{center}
		%
		\label{tab:pattern_distribution}
	\end{center}
\end{table}
  
\end{document}